\documentclass[pre,amsmath]{revtex4}
\usepackage{graphicx}
\usepackage{bm}
\usepackage{latexsym}
\usepackage{pifont}
\usepackage{float}
\usepackage[utf8]{inputenc}

\usepackage{url}
\usepackage{lineno}
\usepackage{microtype}
\usepackage{hyperref}

\begin{document}


\author{Yoav Ravid\,$^{1,*}$}
\author{Samo Peni\v{c}\,$^{2}$}
\author{Yuko Mimori-Kiyosue,$^{3}$}
\author{Shiro Suetsugu,$^{4,5,6}$}
\author{Ale\v{s} Igli\v{c}\,$^{2}$}
\author{Nir S. Gov\,$^{1,*}$}
\affiliation{$^1$Department of Chemical and Biological Physics, Weizmann Institute of Science, Rehovot, Israel \\
$^2$Laboratory of Physics, Faculty of Electrical Engineering, University of Ljubljana, Ljubljana, Slovenia\\
$^3$Laboratory for Molecular and Cellular Dynamics, RIKEN Center for Biosystems Dynamics Research, Minatojima-minaminachi,
Chuo-ku, Kobe, Hyogo 650-0047, Japan\\
$^4$Division of Biological Science, Graduate School of Science and Technology, Nara Institute of Science and Technology
8916-5, Takayama, Ikoma, Nara, 630-0192, Japan\\
$^5$ Data Science Center, Nara Institute of Science and Technology, Ikoma 630-0192, Japan\\
$^6$ Center for Digital Green-innovation, Nara Institute of Science and Technology, Ikoma 630-0192, Japan
}

\title{Theoretical model of membrane protrusions driven by curved active proteins}

\begin{abstract}

Eukaryotic cells intrinsically change their shape, by changing the composition of their membrane and by restructuring their underlying cytoskeleton. We present here further studies and extensions of a minimal physical model, describing a closed vesicle with mobile curved membrane protein complexes. The cytoskeletal forces describe the protrusive force due to actin polymerization which is recruited to the membrane by the curved protein complexes. We characterize the phase diagrams of this model, as function of the magnitude of the active forces, nearest-neighbor protein interactions and the proteins' spontaneous curvature. It was previously shown that this model can explain the formation of lamellipodia-like flat protrusions, and here we explore the regimes where the model can also give rise to filopodia-like tubular protrusions. We extend the simulation with curved components of both convex and concave species, where we find the formation of complex ruffled clusters, as well as internalized invaginations that resemble the process of endocytosis and macropinocytosis. We alter the force model representing the cytoskeleton to simulate the effects of bundled instead of branched structure, resulting in shapes which resemble filopodia.

\end{abstract}

\maketitle

\section{Introduction}

Cells in our body have different shapes depending on their function, and they control their shapes by exerting forces on the flexible plasma membrane \cite{Frey2021}. These forces are mostly due to the underlying cytoskeleton, dominated by the cortical actin network. The actin polymerization near the membrane exerts protrusive forces that can give rise to cellular protrusions, such as filopodia and lamellipodia \cite{Mattila2008}. The control of the actin polymerization in space and time is provided by a host of proteins that nucleate actin polymerization where and when it is needed, and are in turn controlled by different signalling cascades. One mechanism for controlling the spatial pattern of actin polymerization on the membrane, is to couple the actin nucleation to curved membrane components (CMCs), that are both bending locally the membrane and are sensitive to the local membrane curvature (such as BAR domain proteins \cite{Simunovic2015}). This coupling was shown theoretically to give rise to positive and negative feedbacks \cite{gov2018guided}, that can result in pattern formation in both the spatial distribution of the actin nucleators (recruited by the CMCs) and the membrane shape. This coupling between curvature and active protrusive forces was explored for a limited regime of parameters in \cite{Fosnaric2019}. Experimental evidence for this coupling between CMC and protrusive forces has been accumulated in the context of lamellipodia \cite{begemann2019mechanochemical,pipathsouk2021wave} and filopodia \cite{scita2008irsp53,ahmed2010bar,vaggi2011eps8,mancinelli2021dendrite,Feng2022,fox2022cooperative,lee2023lamellipodia} formation.

\begin{figure}
    \centering
    \includegraphics[trim={0.5mm 0.5mm 0.5mm 0.5mm},clip]{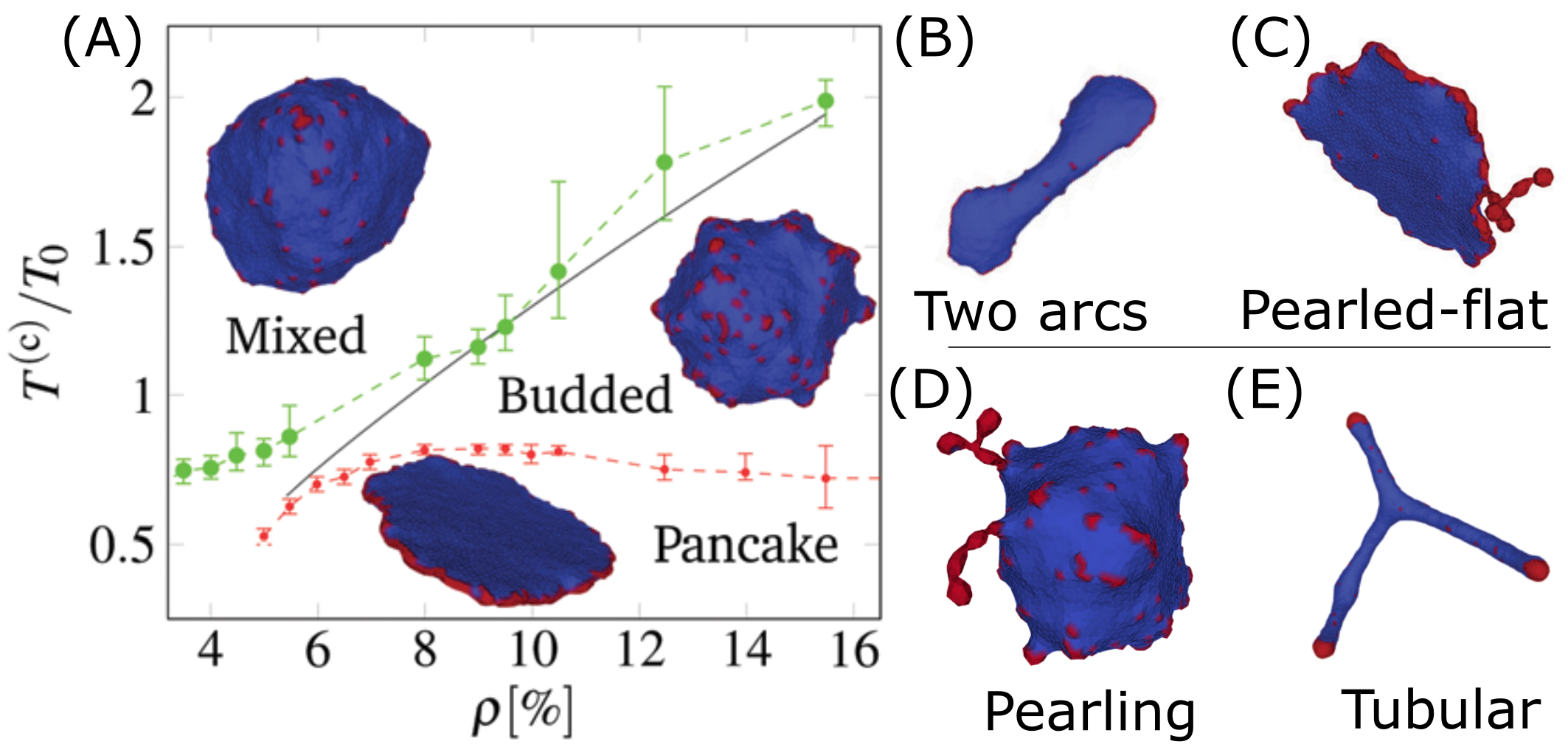}
    \caption{Phases of vesicle shapes driven by curved active CMC, as obtained in \cite{Fosnaric2019}. (A) Phase diagram in the temperature-density plane: mixed (gas), budded, and flattened (pancake). The gas phase is dominated by entropy, hence appears at either high temperatures or low densities. The pancake phase is dominated by having favorable binding and bending energy, where the active forces are all radial and stabilize the flat shape. This phase requires large stable CMC cluster, and so can only appear at low temperatures. The budded phase appears between the two other phases.
    At a CMC density that is lower than the minimal value needed for a closed circular rim, the pancake shape changes to B) a two-arcs phase, while when the CMC concentration is very high the pancake forms pearled extensions that contain the surplus CMC (C). There are two other phases in different regimes: (D) The pearling phase appears at higher CMC density, where most of the CMC aggregate into long necklace-like clusters that minimize the protein-protein binding energy (phase-separation of CMC), and (E) highly-elongated (tubular) phase for flat CMCs, where large CMC caps can exert a strong force that pulls out elongated tubes. Pictures taken from \cite{Fosnaric2019} Figs.4c,7d, and SI.}
    \label{fig:1_Fosnaric_phases}
\end{figure}

A summary of the vesicle shapes that we found in \cite{Fosnaric2019} are shown in Fig.\ref{fig:1_Fosnaric_phases}, explored as function of temperature and CMC density (Fig.\ref{fig:1_Fosnaric_phases}A). The main phases which were identified are \cite{drab2023modeling}:
\begin{itemize}
    \item Diffused CMC-gas phase, where CMC are dispersed as entropy dominates over bending and binding energies.
    \item Budded phase, where binding and bending leads to CMC forming hemispherical clusters at the CMC spontaneous curvature.
    \item Flattened "pancake" phase, where the active forces push the CMC outwards, leading to a large CMC cluster along the rim, with two flat bare membrane disc regions. Low temperature is required to prevent lateral membrane fluctuations and thermal diffusion of the CMC from breaking up the rim cluster.
\end{itemize}
The pancake phase is quite dynamic, and tends to form "ruffles" along the edges. With insufficient density of CMC, there is a "two-arc" phase with multiple flat edges connected by elongated membrane (Fig.\ref{fig:1_Fosnaric_phases}B). If the CMC density if high, the excess CMC form pearled structures along the rim of the pancake (Fig.\ref{fig:1_Fosnaric_phases}C). 

When the active force is weak or zero (passive CMC), at low temperatures the system is phase-separated into energy-minimizing "pearled necklace" of CMC clusters, each at the CMC spontaneous curvature (Fig.\ref{fig:1_Fosnaric_phases}D). When the force is strong and the CMC have low spontaneous curvature (flat), there is a phase of highly elongated "tubular" vesicles, where CMC caps apply large forces that pull membrane tethers (tubular protrusions) (fig.\ref{fig:1_Fosnaric_phases}E).

Here we expand the analysis of the coupling between the spontaneous curvature of the CMC and protrusive forces, by exploring the patterns that form as function of the natural parameters that the cell can manipulate, such as the strength of the actin-driven force, the binding strength between the CMCs and the spontaneous curvature of the CMCs. By gaining a fuller understanding of the space of shapes that this coupling can produce, we are able to explore two more complex configurations: a mixture of two CMCs of different intrinsic curvatures, and CMCs that induce aligned active forces which model the effects of actin bundling\cite{mellor2010role}. These more complex systems, can be compared to important biological phenomena, such as endocytosis\cite{doherty2009mechanisms} and filopodia.

\section{The Model}

We follow the same coarse-grained continuum model used previously \cite{Fosnaric2019} and \cite{Sadhu2021}, where the physics of the cell shape is described by differential geometry and very few energy components \cite{Frey2021,gov2023physical}. The lipid bilayer membrane is modeled as a 2D flexible sheet, with zero spontaneous curvature, except where there are CMCs. Each CMC on the membrane surface represents a complex of proteins that have a specific spontaneous curvature. The energy of the surface is modeled by the Helfrich hamiltonian
\begin{equation}\label{eq:1}
     H_\text{bending}= \iint \; \frac{\kappa}{2} \left(C_1 + C_2 - C_0 \phi \right)^2  
\end{equation}
which penalizes deviation of the shape, given by the local curvatures $C_1$ and $C_2$, from a preferred local shape, determined by the CMC relative lateral density $\phi$ and the CMC's preferred membrane curvature $C_0$.
To simulate, we discretize the system as a closed vesicle described by a graph $V,E$ (vertices and edges respectively) with vertices representing small area patches of either bare lipid bilayer or CMC. Note that the simulation does not have an intrinsic length scale, however the edge length has to represent lengths larger than tens of nanometers for the coarse-grained model to be physically valid. We therefore obtain the following discretized energy
\begin{equation}
    E=\sum_{i\in V} \frac{\kappa}{2}\left(2h(i)-\rho_i C_0\right)^2 A(i) + \sum_{\left\langle i,j\right\rangle \in E} -w \rho_i \rho_j  + \sum_{i \in V} w_{ad} \;\theta\left(z_i-z_0 \right)
    \label{eq:2}
\end{equation}
where $\rho_i=1$ for a CMC vertex and $\rho_i=0$ for a bare vertex, such that the overall density of CMC is given by $\rho=\sum_{i \in V}\rho_i/N$, where $N=4502$ is the total number of vertices in our simulations.
The first term is a discretized version of the bending energy (Eq.\ref{eq:1}), $\kappa$ is the bending modulus, $h(i)$ is the mean curvature calculated at each vertex $h=(C_1+C_2)/2$, $C_0$ is the spontaneous curvature of a CMC, and $A(i)$ is the area assigned to the vertex. The second term is the CMC-CMC nearest-neighbor binding energy, going over the edges $\left\langle i,j\right\rangle$, where $w$ is the binding energy per bond. The third term is adhesion energy of the membrane to a flat rigid surface located at $z=z_0$, which applies to all the nodes that are within a distance of $\ell_{min}$ from this surface. The membrane is prevented from moving below $z_0-\ell_{min}$.

This energy model is used in a Monte-Carlo (MC) simulation \texttt{Trisurf-ng}, described in \cite{Fosnaric2019}, where random movement of vertices and bond flips of edges are accepted if they lower the energy or according to a Boltzmann probability: $P=\exp{\left(-\Delta E-W_i\right)}$ 
where $W_i$ represents the work done by the active forces on each node that contains a CMC, as follows
\begin{equation}
    W_i =  - f \hat{n}(i)\cdot \delta\vec{x}_i 
\end{equation}
where $\hat{n}(i)$ is the local outwards normal unit vector, and  $\delta\vec{x}_i$ is the vertex displacement.

The shift in the locations of the vertices are limited such that the length of each edge remains within this range: $\ell_{min}<\ell<\ell_{max}$. The edge length and adhesion surface constraints are enforced by rejecting any MC moves which violate them. In a passive system this would lead to thermal equilibrium, but the active work term is unbounded from below, so the system is out of equilibrium. The MC simulation does not have time-scale, as it does not include the hydrodynamic flows and dissipative processes that determine the relaxation time-scales of the membrane shape changes. It does allow us to follow the shape dynamics by evolving the system along decreasing energy gradients, so the trajectory in shape space is correctly described.

The parameters in the model, used in this paper, are given in table \ref{tab:simulation_parameters}. All the energies in the model are in units of $k_{B}T$ ($\kappa$, $w$), while the external force $f$ is in units of $k_{B}T/\ell_{min}$.

In addition, we implement optional models of inhibition of the force on the CMC by neighbors, based on \cite{Sitarska2021} which shows different protein species can inhibit the activity of polymerization, inhibiting the actin recruitment and thus force on the CMCs. We implement a proportional inhibition, where an active $\left(1\right)$ and inhibiting $\left(2\right)$ CMC species exist
\begin{equation}
\label{eq:proportional}
    f^{prop}_i = f\frac{1}{N_\text{neighbors}}\sum_{\left\langle i,j\right\rangle} \left(1-\rho_j^{\left(2\right)}\right)
\end{equation}
We also implement a disabling inhibition, where any inhibiting CMC species completely disables the force on it's neighbors.
\begin{equation}\label{eq:complete}
    f^{dis}_i = f\prod_{\left\langle i,j\right\rangle} \left(1-\rho_j^{\left(2\right)}\right)
\end{equation}

In biological filopodia, the actin filament are known to bundle by cross-lining proteins \cite{Faix2006}. Our model does not have a true representation of the cytoskeleton structure, but we can simulate this bundling by adding an alignment to the force on the active CMCs, since the shared internal actin bundle would apply a force in the same direction. This is added as an Vicsek-like interaction \cite{vicsek2012collective}
\begin{equation}
    \hat{f} =  \frac{\hat{n}_i + s \sum_{r} \hat{n}_j } {\left|\hat{n}_i + s \sum_{r} \hat{n}_j \right|}
\end{equation}
The direction of force on CMC vertex $i$ $\hat{f}_i$ is a weighted average of the normal direction plus a contribution from all the vertices $j$ a distance $r$ from the vertex $i$ with a weight of $s$, normalized. This replaces the $\hat{n}(i)$ term in the work term i.e. the unmediated local normal. This is superficially similar to the normal Vicsek model \cite{vicsek2012collective}, where self-propelled particles similarly align their direction with neighbors, producing flocking behavior, but here the CMCs/particles are connected to each other and embedded in a 2D flexible sheet, and we use force in a MC simulation instead of velocity in a Langevin simulation. 

\section{Materials and Methods}

\subsection{Computational Methods}
The simulations were run using trisurf-ng \cite{Fosnaric2019} version fb86a41 ("Modeled\_trisurf" branch) (see \ref{section:data availability}) with a tape file modified from the available default with the different physical parameters (see \ref{tab:simulation_parameters}), and additional simulation running parameters of nshell=30, mcsweeps=50,000-200,000, iterations=100-1,000 (depending on the desired time resolution). Each simulation with a set of parameters was ran independently ("embarrassingly parallel"), which took about two weeks to finish, with occasional restarts and expansion of the space limits (nxmax). The resulting VTU files were viewed and colored in ParaView, but further analysis and graph generation were done by separate python scripts.

\subsection{Experimental Methods} 
The cell culture and lattice light sheet microscopic observation U-251 cells were obtained from the Japanese Collection of Research Bioresources Cell Bank. The IRSp53 knockout (KO) cells were generated by the CRISPR/Cas9 system, as described previously \cite{mashiko2013generation}. The guide RNA targeting the first exon of IRSp53 (CCATGGCGATGAAGTTCCGG) was designed using the server http://crispr.mit.edu and inserted into the pX330 vector \cite{mashiko2013generation}. After transfection, the cells were cloned by monitoring the GFP fluorescence from the reporter plasmid pCAG-EGxxFP with the IRSp53 genome fragment using a fluorescence-activated cell sorter [FACSAria (BD)] \cite{hanawa2019phagocytosis}. The expression of GFP or GFP-IRSp53 in the IRSp53 knockout cells was performed by the retrovirus-mediated gene transfer, as described previously \cite{hanawa2019phagocytosis}. All cell lines were cultured in high glucose DMEM (Thermo Fisher Scientific) supplemented with $10\%$ bovine calf serum (Thermo Fischer Scientific) and $1\%$ penicillin-streptomycin solution (Thermo Fischer Scientific) and stored in an incubator at $37^{o} C $ in $5\%$ $\text{CO}_\text{2}$ and humidified conditions. The cells were seeded on coverslips and then imaged with the Lattice light-sheet microscope built in the Mimori-Kiyosue laboratory at RIKEN Center for Biosystems Dynamics Research following the design of the Betzig laboratory \cite{chen2014lattice} as described previously \cite{nishimura2021filopodium}.

\section{Force-binding strength phase diagram}

In \cite{Fosnaric2019} the phases of the vesicle with active CMC, were mostly explored as function of temperature and global density of CMC. However, the cell can more easily modify other parameters, such as the strength of the protrusive forces produced by actin polymerization and the binding strength between neighboring CMC. The rate of actin polymerization recruited to the CMC can be controlled by the cell through various proteins \cite{welch2002cellular,carlier2007control,campellone2010nucleator}. The effective binding strength between the neighboring CMC can similarly depend on the lateral concentration and character of the proteins that form the CMC \cite{pipathsouk2021wave}, as well as on the type of lipids between the CMC \cite{bohinc2003interaction}. The cell can modify these internal parameters spontaneously or in response to external signals.

\begin{figure}
    \centering
    \includegraphics[width=0.90\linewidth]{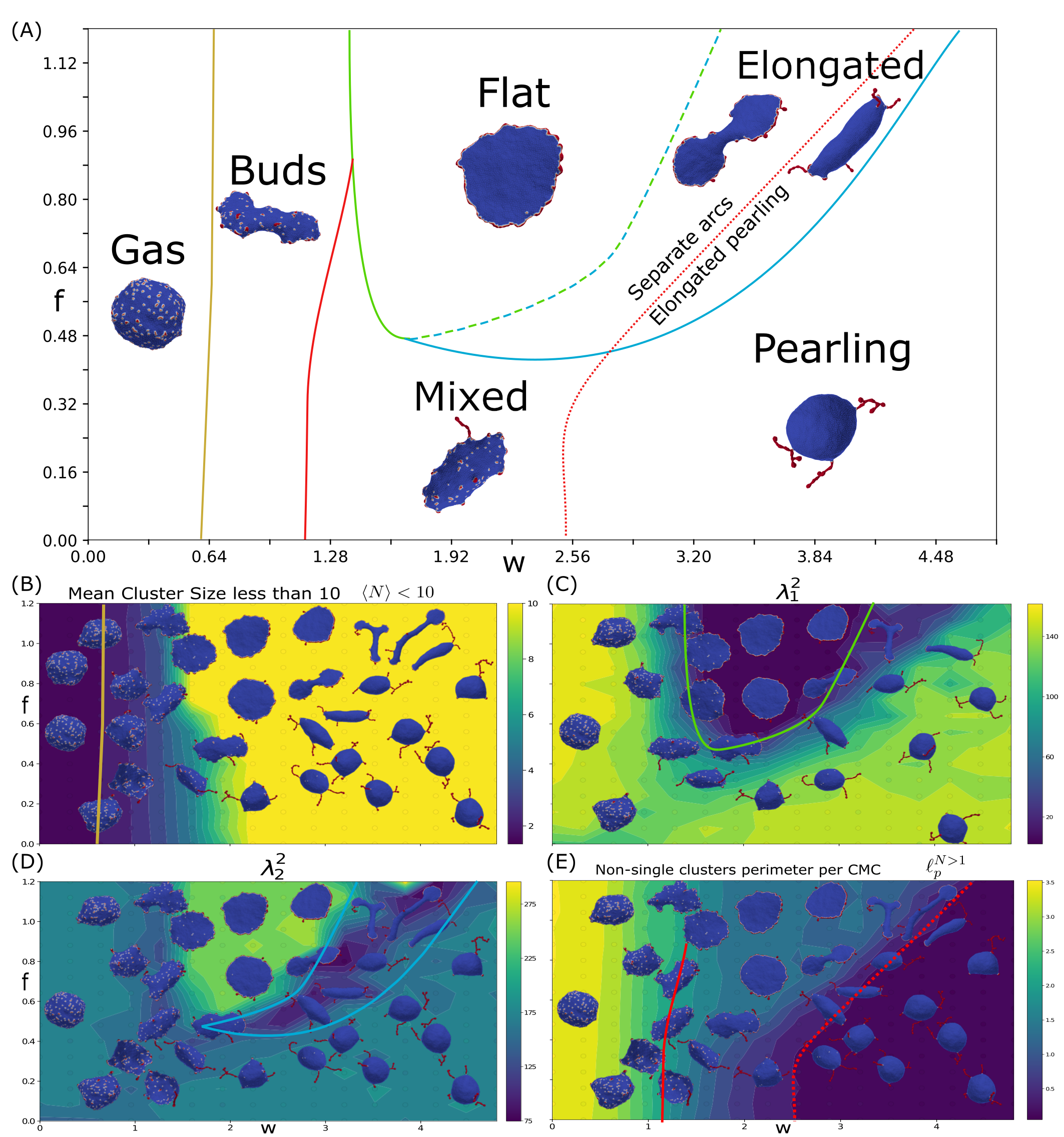}
    \caption{Force-binding strength plane. (A) Phase diagram as function of $f$ and $w$, with: $\kappa=20$, $C_0=1$, and $\rho=10\%$. The different phases are indicated by their names, and a typical snapshot of the vesicle after a long simulation is shown. The transition lines between the phases were drawn according to the measures shown in the bottom panels. The gas and buds phase is separated by mean cluster size $\left\langle N\right\rangle=1.5$ (yellow solid line), as obtained from (B). The green line denotes the boundary of the flat phase, obtained approximately from a contour of the first (small) gyration eigenvalue $\lambda_1^2$, which is minimal for flat shapes (C). The light blue line denotes the boundary of the elongated shapes, roughly following a contour of the second (intermediate) gyration eigenvalue $\lambda_2^2$ (D). The transition line between the buds and mixed phases is given by a contour of CMC perimeter length ($\bar{\ell}_p=1.875$, red solid line), extracted from (E). Finally, the pearling phase transition line (red dotted line) is drawn along the contour of small CMC perimeter length ($\bar{\ell}_p=0.375$), from (E). In panels (B-E) we plot heatmaps of the following quantities: (B) Mean cluster size for clusters smaller than 10, $\left\langle N\right\rangle >10$ (C) first (small) gyration eigenvalue $\lambda_1^2$, (D) second (intermediate) gyration eigenvalue $\lambda_2^2$, (E) Mean CMC cluster perimeter length (excluding isolated CMC) $\bar{\ell}$.}
    \label{fig:2_FW_Phase_Diagram}
\end{figure}

We scan over the force $f$ and binding strength $w$ parameters plane (Fig.\ref{fig:2_FW_Phase_Diagram}A), with the other parameters of the model having the following constant values: The bending modulus is taken to be $\kappa=20 K_{B}T$, which is a typical value for lipid bilayers. The spontaneous curvature of the CMC is taken to be $C_0=1\ell_{min}^{-1}$, representing highly curved objects on the membrane. The CMC density is $\rho = 10\%$, which is sufficient to form the pancake shapes that require a complete circular cluster of CMC along the vesicle rim \cite{Fosnaric2019}. 

We find that the simulated vesicles can be divided into several distinct phases: gas phase, budded phase, pancake phase, and pearling phase. In addition there are more ambiguous, and possibly transient, elongated and mixed phases (Fig.\ref{fig:2_FW_Phase_Diagram}A). In order to distinguish between these phases, we use four measures that characterize the vesicle shape and the CMC cluster organization:
\begin{itemize}
    \item Mean cluster size $\left\langle N \right\rangle$
    \item $1^\text{st}$ eigenvalue of the Gyration tensor $\lambda_1^2$
    \item $2^\text{nd}$ eigenvalue of the Gyration tensor $\lambda_2^2$
    \item Length of CMC-bare membrane boundary $\ell_p$
\end{itemize}

The mean cluster size is averaged over all the CMC clusters, each cluster $i$ having a size $ N_i$ of vertices
$$ \left\langle N \right\rangle = \frac{\sum_i N_i}{\sum_i 1}=\frac{N_\text{vertex}}{N_\text{clusters}}$$
We plot this measure (Fig.\ref{fig:2_FW_Phase_Diagram}B), extracted after the simulation reaches its steady-state regime, where the measures do not change on average (see SI). we see that it allows to clearly distinguish the gas phase, which has small cluster sizes (yellow line in Fig.\ref{fig:2_FW_Phase_Diagram}A denotes $\left\langle N \right\rangle=1.5$). However, it is rather poor at separating the condensed phases, which all have large clusters but differ greatly in their morphology and cluster organization. This is due to the dependence of this measure on the number of clusters, which gives large weight to small single-vertex clusters. This makes this measure too noisy to distinguish between the other phases, except for the gas phase which mostly contains single-vertex clusters.

We therefore use morphological measures in order to clearly distinguish between the different phases where the CMCs are condensed in large clusters. The morphology of the vesicle is quantified by the eigenvalues of the gyration tensor $\lambda_i^2$. The gyration tensor \cite{theodorou1985shape} is defined as the average over all the vertices, with respect to the center of mass (similar to the moment of inertia tensor for equal-mass vertices)
$$ R_{G\;ij} = \left\langle r_i r_j \right\rangle=\frac{1}{N}\sum_{vertices} \left(\begin{matrix} x^2 & xy & xz \\ xy & y^2 & yz \\ zx & yz & z^2 \end{matrix}\right)$$
This can be visualized by a unique ellipsoid which has the same gyration tensor
$$ \boldsymbol{x}^T \, \boldsymbol{R_G}^{-1} \, \boldsymbol{x}=\frac{\left(\boldsymbol{x}\cdot\boldsymbol{e_1}\right)^2}{\lambda_1^2}+\frac{\left(\boldsymbol{x}\cdot\boldsymbol{e_2}\right)^2}{\lambda_2^2}+\frac{\left(\boldsymbol{x}\cdot\boldsymbol{e_3}\right)^2}{\lambda_1^2}=3 $$
The eigenvectors $\boldsymbol{e_i}$ of the gyration tensor are the directions of the semi-axes of the equivalent ellipsoid and the eigenvalues are their length squared divided by $3$, ordered by their size: $\lambda_1^2\le\lambda_2^2\le\lambda_3^2$. The first eigenvalue $\lambda_1$ essentially gives how thin is the ellipsoid, and is low for both pancake and highly elongated (linear) shapes. The second eigenvalue $\lambda_2$ is large for the pancake shape (as it is roughly equal to the largest eigenvalue $\lambda_2\sim\lambda_3$), but is minimized for elongated shapes, where it similar to the value of the smallest eigenvalue, $\lambda_2\sim\lambda_1$. In Fig.\ref{fig:2_FW_Phase_Diagram}C,d we plot the eigenvalues $\lambda_1^2,\lambda_2^2$, respectively. We find that the phase of pancake shapes is distinguished by the lowest $\lambda_1^2$ (green and dashed green-light blue lines in Fig.\ref{fig:2_FW_Phase_Diagram}A), indicating its flatness.

We identify a new phase of elongated shapes, which is distinguished by the lowest values of $\lambda_2^2$ (between the light blue and dashed green-light blue lines in Fig.\ref{fig:2_FW_Phase_Diagram}A). These elongated phases are somewhat similar to the "two-arc" phase found in \cite{Fosnaric2019}, which appeared when there are not enough CMCs to form a complete circular cluster along the flat vesicle rim. However, here we do have enough CMC to form a complete circular cluster, as shown in the "flat" regime. The origin of the elongated shapes as $w$ increases beyond the "flat" phase is due to the formation of transient or stable pearling clusters. 
These cluster effectively sequester enough CMC to prevent the formation of the complete circular cluster, leading to two curved regions that collect the CMC and stretch the vesicle due to the active forces. The CMC clusters have the shape of flat arcs near the boundary with the "flat" phase, while closer to the "pearling" phase the clusters are pearled and localized near the curved tips of the vesicle.

While the "core" of the phases distinguished by $\lambda_{1,2}^2$ is clear, the edges are much less sharp, due to lack of statistics, long evolution time, and the fact that intermediate shapes do exist. There is also no obvious normalization: The volume changes greatly, and the area is only approximately conserved. For our $N_{vertex} = 4{,}502$ The flat phase is found around $\lambda_1^2<50$, and the elongated phases is found around $80<\lambda_2^2<150$.

Finally, we wish to distinguish the phases where the CMCs form pearled clusters. The most outstanding property of the pearled clusters is that they phase-separate between the CMC and the bare membrane, as also predicted within the theory of self-assembly \cite{Fosnaric2019}. We therefore measure the average length of the CMC-bare membrane boundary $\bar{\ell}_p$, per CMC, for all clusters larger than $1$ (see SI section 1, Fig.S1)
$$ \bar{\ell}_p= \left\langle \frac{\ell_{p i}}{N_i}\right\rangle_{N_i>1}$$
The phase with pearling clusters is distinguished by having very low $\bar{\ell}_p<0.375$ (Fig.\ref{fig:2_FW_Phase_Diagram}E). We find that this measure identifies the pearled clusters both in the pearling and in the elongated phases (red dotted line in Fig.\ref{fig:2_FW_Phase_Diagram}A). In addition, a contour of this measure allows us to separate the mixed phase, where the CMC are in both buds and pearled clusters, from the phase that contains only buds (red solid line $\bar{\ell}_p<=1.875$ in Fig.\ref{fig:2_FW_Phase_Diagram}A,E).

Note that we do not know if these phases are necessarily the absolute steady-states of the system in the limit of infinite time. The system might be trapped in a local meta-stable configuration due to dynamical barriers that would require unreasonably long simulations for them to escape. For example, in the regime of low force $f$ and large binding strength $w$, the global minimum energy configuration should have all the CMC in a single pearled cluster, but during the merging of the pearled clusters into a single cluster they have to overcome bending energy barriers that hinder this process \cite{Golani2019}. In other regimes, such as the elongated phase, we do not know if a stationary steady-state even exists, since the presence of active forces may induce a constantly changing configurations. In the SI section 2 we give a simple analytic calculation that gives reasonably well the transition line between the pearled and flat phases, which are the main stable condensed phases in this phase diagram (Figs.S2,S3). 

\begin{figure}
    \centering
    \includegraphics[trim={0.5mm 0.5mm 0.5mm 0.5mm},clip]{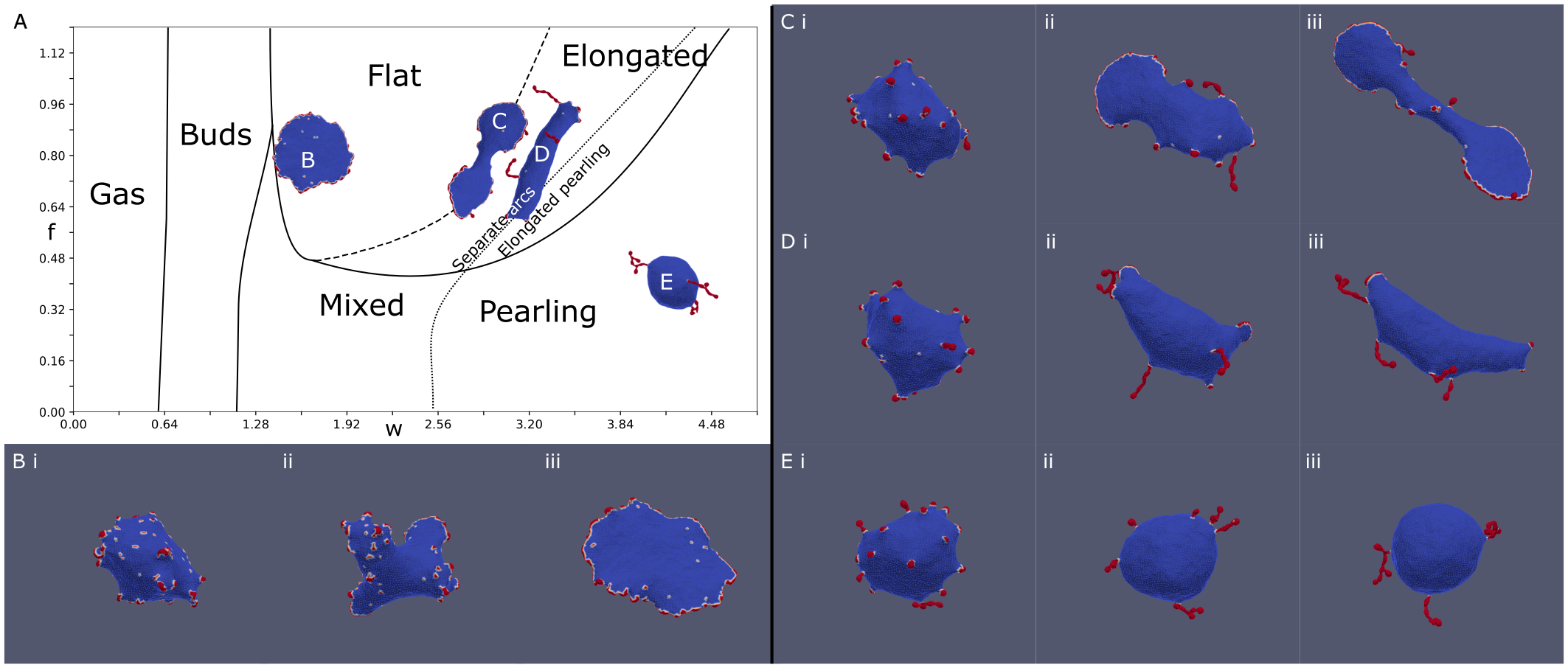}
    \caption{Evolution of the MC simulation at four different points (B-E) denoted on the phase diagram (A) (Fig.\ref{fig:2_FW_Phase_Diagram}A). (B): f=0.8, w=1.6, (C): f=0.8, w=2.88, (D): f=0.8, w=3.20, and (E): f=0.4, w=4.16. The MC time-steps shown in the snapshots are: (i) 10, (ii) 50 (ii) and (iii) 200, and the final time-step (299) is shown on the phase diagram (A). At time (i), all simulations are in the budded state. At time (ii), arc and pearling clusters begin to form, favoring arcs for large forces and pearling for large binding strength. At time (iii), the vesicles are close to their final steady-state shapes. The flat simulation (B) generates several arcs in stage (ii), which coalesce to form a circular stable rim. The pearling simulation (E)  generates pearling clusters (ii) which coalesce into a few larger clusters (coarsening). In contrast, the elongated simulations generate both arcs and pearled clusters at the intermediate stage (ii). These arc-like clusters are sufficient stretch the vesicle, even in (D), to give rise to the final elongated phase.}
    \label{fig:3_FW_progression}
\end{figure}

The evolution of a handful of chosen simulations are shown in Fig.\ref{fig:3_FW_progression}, showing flat, elongated-flat, elongated-pearling, and pearling phases. All the simulations begin in a disordered uniform distribution of the CMC on the spherical vesicle, but in all of them we find that buds form rather quickly (Fig.\ref{fig:3_FW_progression}B(i)-E(i)). In the budded phase this configuration simply remains stable and does not evolve significantly. It takes longer time for the larger clusters of the flat rim, arcs and pearls to form. The transition lines separating two different vesicle phases, obtained from our simulations, are not precise, and one can obtain either one of the vesicle shapes close to these lines (Fig.\ref{fig:3_FW_progression}A).

To conclude, by exploring the $f-w$ phase diagram, we demonstrate the competition between the protein binding which drives the formation of pearled clusters, and the active force that drives the formation of arc-like clusters at the edge of flat protrusion. This competition is highlighted in the new phases of vesicle morphologies that we found, namely the elongated two-arcs and the elongated-pearled phases. The pearling phase appears for large enough values of $w$, as follows also from the theory of self-assembly \cite{Fosnaric2019}.

\section{Force-spontaneous curvature phase diagram}

We now proceed to explore the interplay between the active force and the spontaneous curvature of the CMC in determining the morphology of the vesicle. We chose the parameters for a new set of simulations such that we are in the flat phase when the CMC are highly curved: $\rho=20\%$, $\kappa=28.5$, $w=2$. The resulting phase diagram is shown in Fig\ref{fig:4_cf_and_coral}A.

We find several phases: budded phase, flat phase, elongated (arcs) phase and highly-elongated (tubes) phase. Here the boundaries between the different phases were drawn by eye, due to relative sparse scan over the parameters, and the self-evident boundaries (Fig.\ref{fig:4_cf_and_coral}A). In this parameter regime, we do not find any pearled phase, with the budded phase remaining stable due to the bending energy barrier that prevents buds merging (note that the bending modulus is larger here), and lower relative $w$. Similar to the force-binding strength system (Fig.\ref{fig:2_FW_Phase_Diagram}A), where the budded and pearled phases exist for low active force, we also find that as the active force is increased the budded phase is destabilized to form the flat phase (Fig.\ref{fig:4_cf_and_coral}A).

The flat phase is destabilized as the spontaneous curvature decreases due to the following mechanism: as $C_0$ decreases the thickness of the rim cluster increases, which means that there are not enough CMC to complete a circular cluster around the edge of the flat shape. The morphology then changes into local arc-like clusters that pull the vesicle into elongated shapes. The elongation of these vesicles depends on the magnitude of the active force.

The main feature of this phase diagram is the appearance of the highly-elongated tubular phase, where the entire vesicle is stretch into a several tubes that are pulled by CMC clusters at their tips. We can theoretically estimate the location of the phase transition line, above which a vesicle will become highly-elongated, by comparing the force exerted by the active CMC cluster and the restoring force of the emerging membrane tube due to bending (Fig.\ref{fig:4_cf_and_coral}B). A hemispherical CMC cap with radius $r=2/C_0$ minimizes the bending energy (Eq.\ref{eq:1}): $E\propto\left(\frac{1}{r_1}+\frac{1}{r_2}-C_0\right)^2$, and maximizes the pulling force (since adding any more CMCs to the cluster, beyond the hemisphere, adds force in the opposite direction). The total pulling force of this hemispherical cluster is given by
\begin{equation} F_{pull} = f \cdot\underbrace{\frac{1}{2}}_{geometry}\cdot\underbrace{\frac{2\pi(2/C_0)^2}{s_0}}_{\#vertices}\label{eq:Fpull}
\end{equation}
where $s_0$ is the area per vertex, and $2\pi r^2/s_0$ is the number of CMC in the cluster.
This hemispherical cap pulls a tube with the same radius from the main vesicle body. Note the extra factor of $1/2$ due to the hemispherical shape of the cup, compared to the calculation done for a flat cluster of active proteins in \cite{Fosnaric2019}.

Assuming the restoring force is dominated by the bending energy of the membrane tube, it is given by (Eq.\ref{eq:1}) \cite{Fosnaric2019}
\begin{equation} F_{restore}=\frac{\kappa}{2}\frac{2\pi}{(2/C_0)}\label{eq:Frest}
\end{equation}
The highly elongated shape is initiated when the pulling force is greater than this restoring force, so the critical value is given by equating Eqs.\ref{eq:Fpull},\ref{eq:Frest}, which gives
\begin{equation}\label{eq:force_balance}
f = A C_0^{3}
\end{equation}
where $A$ is a constant determined by the constant parameters of the simulation (bending modulus and average area per vertex). Plotting this simple cubic relation in Fig.\ref{fig:4_cf_and_coral}A (blue solid line, where we fit the value of $A$), shows a good agreement with the observed boundary of the regime of the highly-elongated tubular shapes on the phase diagram. Note however that the shapes of the vesicles at the transition to the tubular phase are not always simple cylindrical tubes with hemispherical clusters at their tips (Fig.\ref{fig:4_cf_and_coral}A), as the analytic model assumes (Fig.\ref{fig:4_cf_and_coral}B).

To conclude this section, we have shown that active CMC give rise to flat protrusions when they are highly curved. Tubular protrusions can form for weakly curved active CMC, while for highly curved CMC the active force needed to produce such slender protrusions increases extremely fast. In the next sections we explore how slender tubular protrusions can be produced with highly curved active proteins, by either changing the effective curvature of the CMC cluster, or by increasing the effective pulling force of the cluster.

\section{Multiple curvature}

Real cells have many species of membrane protein of both convex and concave intrinsic curvature. While these membrane proteins have distinct curvatures, the effective curvature of a cluster of CMC may depend on the composition of the cluster, if it contains CMC of different spontaneous curvatures. In order to form clusters of mixed curvatures, we explore vesicles that contain CMC of different curvatures (concave and convex), that bind to each other equally. If the two CMC types bind only to their own kind, they form separate clusters on the vesicle, and their coupling with each other due to curvature alone is rather weak (see SI). The convex CMC maintain their activity, as in the previous sections, while the concave CMC is passive.

\begin{figure}
    \centering
    \includegraphics[trim={0.5mm 0.5mm 0.5mm 0.5mm},clip]{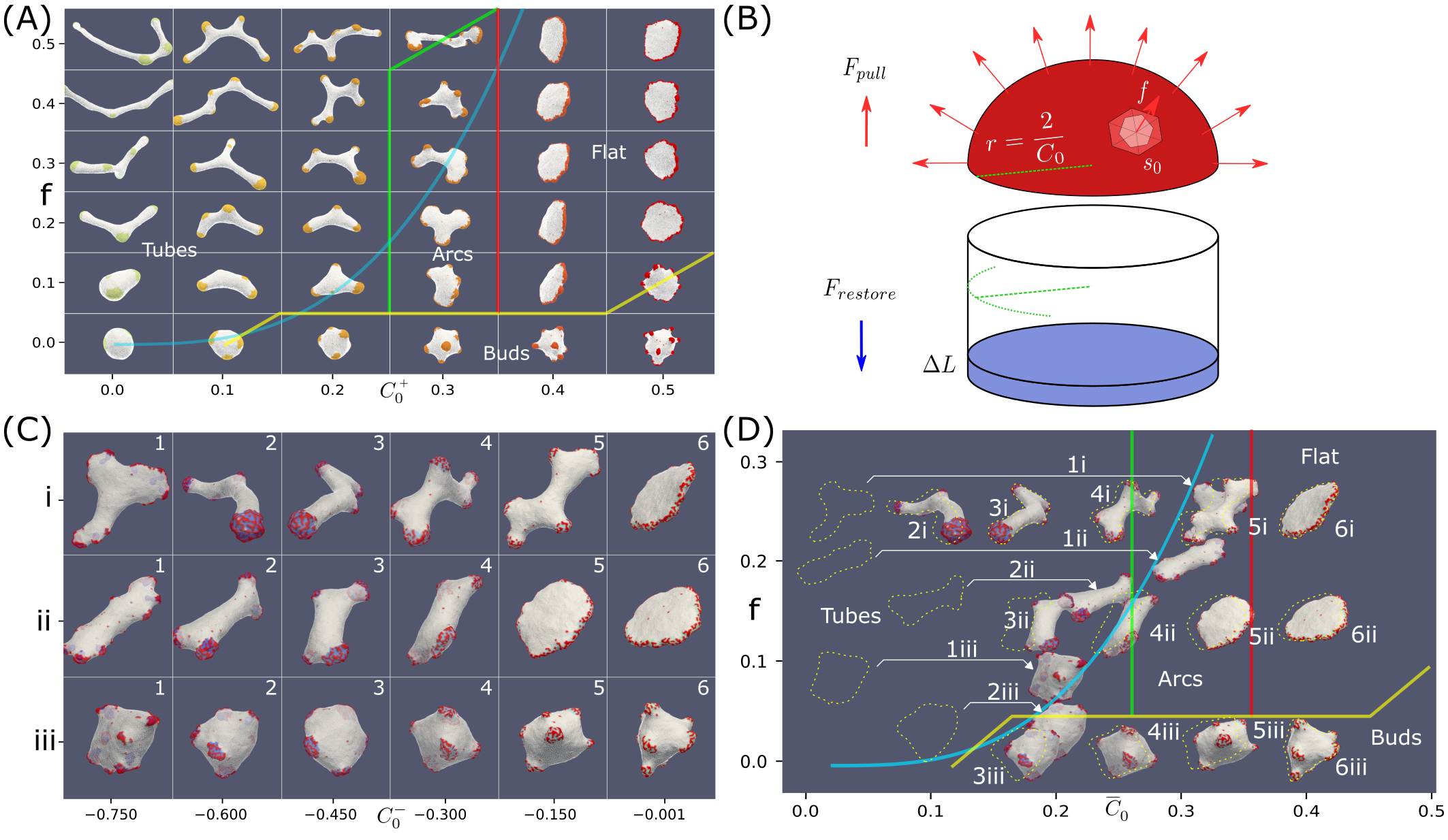}
    \caption{(A) Phase diagram in the force-spontaneous curvature plane, using the parameters: $\rho=20\%$, $\kappa=28.5$, $w=2$. The different phases are denoted by their typical shapes, and the thin colored transition lines were drawn by hand (yellow, red and green). With no or weak force, we find a budded phase. As the force is increased, we find for the high spontaneous curvature the flat phase. As the spontaneous curvature is reduced, the flat phase is observed to give way to an "arcs" phase, which is finally replaced by a highly-elongated tubular phase. The thick blue line denotes the theoretical calculation for the transition line that bounds the tubes phase, which is a cubic equation: $f=A C_0^3$ (Eq.\ref{eq:force_balance}), where we use: $A\approx10.6$. This equation is derived from the force balance shown schematically in (B). (C) Typical steady-state snapshots of simulations with a mixture of CMC: active convex CMC ($C_0=0.8$, $f=0.5$, $\rho=10\%$), and passive concave CMCs ($\rho=10\%$) with different concave curvatures $C_0^-$ (along the x-axis). We show here three cases: i) no inhibition of the active convex CMC, ii) proportional inhibition, where the force exerted by a convex CMC is proportional to number of non-concave neighbors, and iii) disabling interaction, where the convex CMC do not exert any force if they have a concave neighbor. (D) Mapping of the vesicles shown in (C) to their respective locations in the force-spontaneous curvature phase diagram (A), using the average force and spontaneous curvature of the mixture (dashed outlines). The snapshots are shown at shifted locations, according to the effective curvature when we take into account the phase-separation of the concave CMC, into internalized sacks. These shifts in locations are most dramatic for 1i,1ii,2ii,1iii,2iii (indicated by arrows), which places the vesicles in a phase which is appropriate for their shapes.}
    \label{fig:4_cf_and_coral} 
\end{figure}

In Fig.\ref{fig:4_cf_and_coral}C(i) we show snapshots of the steady-state shapes of the vesicles that contain $10\%$ passive concave CMC, i.e. a CMC species with $C_0^{-}<0$ and $f^{-}=0$, in addition to  convex CMCs ($\rho^+ = 10\%$, $f=0.5$, and $C_0^+=0.8$). Both types of CMC have the same binding strength $w=2$, which binds both types equally, leading to strong mixing of the two CMC types. For weakly curved concave CMC ($C_0^{-}=-0.001$) the flat phase remains stable (Fig.\ref{fig:4_cf_and_coral}C(i6)), driven by the convex active CMC. As the concave CMC become more curved (Fig.\ref{fig:4_cf_and_coral}C(i) from right to left) the circular cluster at the rim of the flat shape breaks up, and highly elongated shapes appear (Fig.\ref{fig:4_cf_and_coral}C(i2,i3)).

\begin{figure}
    \centering
    \includegraphics[trim={0.5mm 0.5mm 0.5mm 0.5mm},clip]{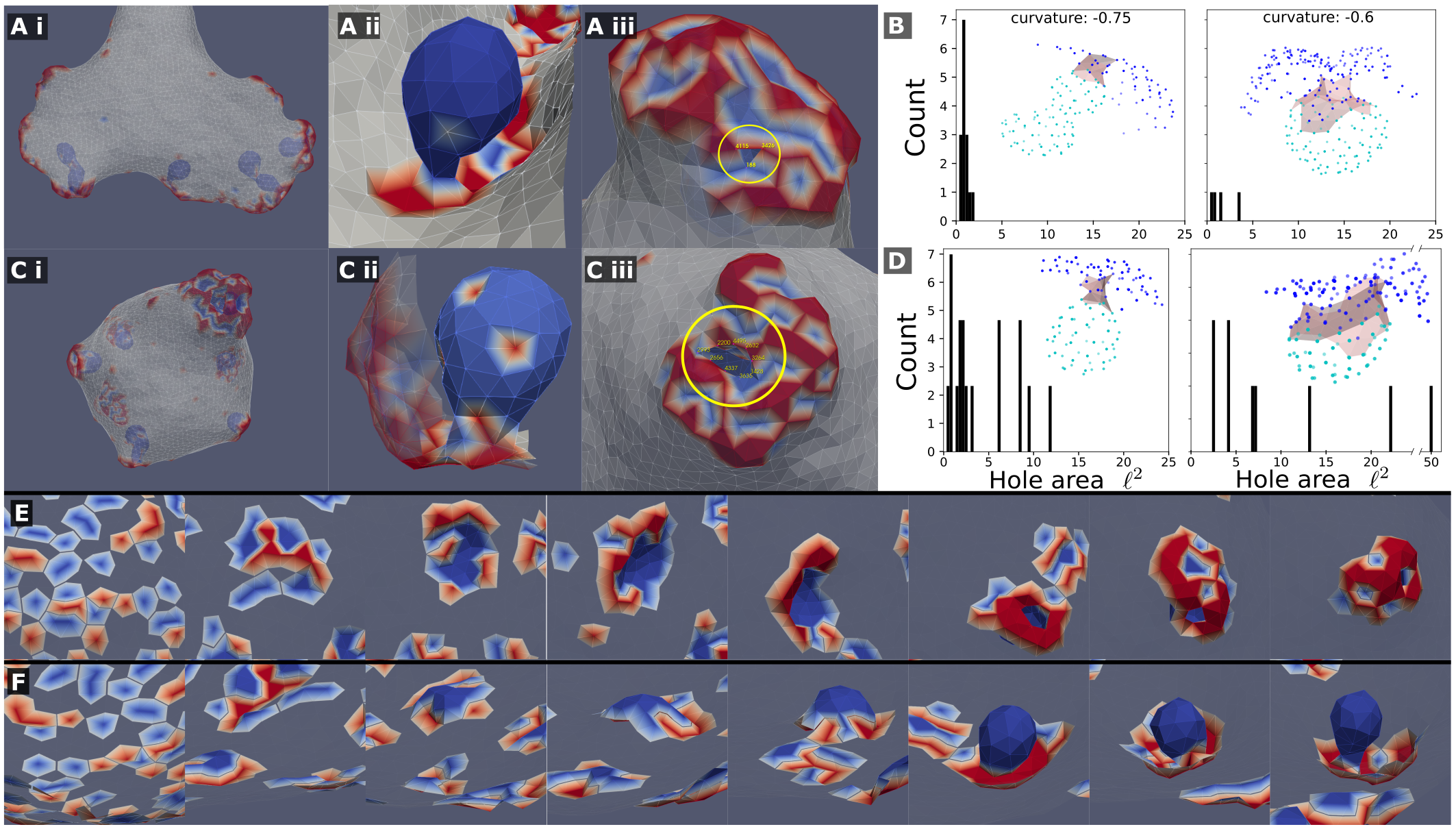}
    \caption{Mixed clusters can precipitate internal sacks, which are composed almost entirely of the concave (passive) CMC, when the concave CMCs are highly curved $C_0=-0.75, -0.6$. This is shown in A(i,ii),C(i,ii) for a system with and without active force, respectively. This internal sack is connected to the outside by a thin neck, or "hole", shown in A(iii) and C(iii). The cross-sectional area of the hole was measured by computing the area of the polygon made from the hole edge, which was picked by hand (vertices). A histogram of the simulated hole sizes is shown for the system with and without active force respectively (B,D). It is clear that the hole size is smaller in systems with force (B), such that it is in the limit of the simulation resolution. The holes are also larger as the spontaneous curvature of the passive concave CMC is smaller. The insets of B,D show typical examples of sacks (light blue nodes) connected to the outer part of the cluster (blue nodes) through the neck region (grey shading). (E) and (F): Snapshots showing the formation of a sack for the system with active force (A), from the initial random state. In (E) we show the cluster viewed from outside of the vesicle (where the bare membrane is rendered invisible), looking down on the patch that forms the sack, while in (F) we show the same process viewed from within the vesicle, where we see clearly the final invagination.}
    \label{fig:5_Coral_sack_shapes} 
\end{figure}

These shapes can be explained by mapping the vesicles in Fig.\ref{fig:4_cf_and_coral}C(i) on the phase diagram (Fig.\ref{fig:4_cf_and_coral}A). For each simulation, we calculate the average spontaneous curvature of the CMC clusters: $C_{0,eff}=\left(C_0^++C_0^-\right)/2$, as well as the average pulling force per CMC: $f_{eff}=f/2$. In Fig.\ref{fig:4_cf_and_coral}D we plot the typical dashed outline of the vesicles from Fig.\ref{fig:4_cf_and_coral}C(i) on the phase diagram according to these effective parameters $C_{0,eff},f_{eff}$. Most vesicles match the shape of the phase to which they are mapped in this way. The only exception is the vesicle with the most concave CMCs (and effective $C_{0,eff}=0$), which is not in the shape of highly-elongated tubes, as suggested by the calculated average parameters, but fits better the arcs phase. This phenomena is due to the concave CMCs phase-separating into internal "sacks" of concave-enriched clusters (Fig. \ref{fig:5_Coral_sack_shapes}Ai), which results in an effective removal of these concave CMC from determining the outer shape of the vesicle. To take this into account, we calculate the effective mean curvature of the CMCs while removing the concave CMC that are contained in the internalized sacks. This is done by including in the calculation of the average curvature only concave CMCs which are connected to at least one convex CMC. Using this revised average spontaneous curvature, we plot the locations of the vesicles on the phase diagram (full snapshots), and find that except for the most curved concave CMC (A1), the locations of the other vesicles is minimally affected. For the case A1, we find that indeed the formations of large sacks of concave CMC, push the vesicle into the arcs regime, compatible with its revised location on the phase diagram. The phase separation of the passive concave CMC into sacks is driven by the minimization of the total bending energy. The highly elongated tubes cost a high bending energy of the bare membrane: in Fig.\ref{fig:4_cf_and_coral}C(i2) the average bending energy of the bare membrane is $\sim25 K_B T$, while in the flatter shapes after the phase separation (Fig.\ref{fig:4_cf_and_coral}C(i1)) the average bending energy of the bare membrane drops to $\sim17 K_B T$.

In addition to the overall vesicle shape in the system of mixed curvatures, we are interested in the character of the CMC clusters. We find that concave and convex CMCs create complex mixed clusters with a "coral"- or "sponge"-like texture (Fig.\ref{fig:4_cf_and_coral}C and close up in Fig. \ref{fig:5_Coral_sack_shapes}Aii). The texture of these clusters seems similar to the membrane ruffles observed in \cite{Sitarska2021} behind the leading edge of motile cells. In this work, the ruffles were attributed to the interaction between concave and convex membrane proteins, that are also involved in the recruitment of the actin polymerization. It was furthermore proposed in \cite{Sitarska2021} that the pattern of ruffles observed in these cells is determined by the interaction between a concave membrane protein that inhibits the actin polymerization, which is recruited by the convex CMC. Motivated by this proposed mechanism, we explored the resulting shapes of the vesicle and CMC clusters when the concave CMCs inhibit the active force exerted by the convex CMCs. We tested two possibilities: inhibition that is proportional to the number of concave neighbors (Eq.\ref{eq:proportional}, Fig. \ref{fig:4_cf_and_coral}C(ii)), and full inhibition with even one concave neighbor (Eq.\ref{eq:complete}, Fig. \ref{fig:4_cf_and_coral}C(iii)). In both cases we find that the effective force is reduced, and that the resulting shapes correspond very well to their locations on the phase diagram (Fig.\ref{fig:4_cf_and_coral}D). The shapes obtained for full inhibition (Fig. \ref{fig:4_cf_and_coral}C(iii)) are very similar to those for a vesicle with a mixture of passive CMC (see SI section 3, Fig.S4). 
Regarding the comparison with the experiments \cite{Sitarska2021}, we conclude from the model that the ruffle texture of the CMC clusters does not crucially depend on the inhibitory interaction between the two CMC types, but rather on their spontaneous curvatures and binding interaction.

Let us now focus on the phase-separated sacks of highly curved concave CMC, which form within the mixed clusters (Fig.\ref{fig:5_Coral_sack_shapes}). We observed that the neck that connects the sacks to the outer part of the cluster is much narrower when the convex CMC exert outwards protrusive forces (compare Fig.\ref{fig:5_Coral_sack_shapes}(Aii) and (Cii)). We quantified the area of the narrowest part of the neck in Fig.\ref{fig:5_Coral_sack_shapes}B,D for the active and passive convex CMC, respectively. The necks are naturally narrower for more highly curved concave CMC. The active convex CMC, which push the membrane outwards, exert an effective pressure force that squeezes the neck into a narrower radius. Note that for the narrowest necks, we are clearly at the limit of the spatial resolution of the simulation. We do not allow membrane fission, and therefore can not describe the process of detachment of such sacks as internalized vesicles \cite{penivc2020budding}, as occurs in cells during endocytosis and macropinocytosis \cite{kay2021macropinocytosis}.

In Fig.\ref{fig:5_Coral_sack_shapes}E,F we show the dynamics of the cluster formation, whereby a patch of passive concave CMC (blues) increase in size, while its rim is populated by active convex CMC (red). In these images the surrounding bare membrane is rendered to be invisible. These simulated dynamics resemble those calculated by another model of macropinocytic cups \cite{saito2021three}, which was based on reaction-diffusion dynamics coupled to active forces.

Finally, when the two CMC types bind exclusively to their own kind, they form separate clusters, with very limited coupling between them (see SI section 4, Fig.S5).

\section{Force alignment}

As we show in Fig.\ref{fig:4_cf_and_coral}A, when the highly curved CMC induce a protrusive force that is directed at the outwards normal, we require an extremely large force in order for the highly elongated tubes to form. However, cells initiate slender, tube-like filopodia protrusions using highly curved membrane proteins, such as IRsp53 \cite{scita2008irsp53,ahmed2010bar,vaggi2011eps8,Feng2022,fox2022cooperative,lee2023lamellipodia}, in agreement with theoretical calculations \cite{mesarec2016closed}.
Within the slender filopodia in cells, the actin filaments are organized into a cross-linked bundle, which efficiently directs the forces of all the polymerizing actin filaments along the protrusion's axis. The actin nucleators at the tip of the filopodia are different from those at the leading edge of the flat lamellipodia \cite{Faix2006,faix2022ena,lee2023lamellipodia}, and initiate the growth of parallel actin filaments that form the bundle at the filopodia core. In our model, since we do not explicitly describe the actin filaments organization, we can only describe the effects of the bundling on the forces exerted on the membrane. To simulate this kind of bundling, we add an alignment term of a Vicsek-like interaction \cite{vicsek2012collective}, which aligns the forces exerted on the membrane by each CMC that is bound in a cluster
\begin{equation}\label{eq:vicsek_force}
    \hat{f}_i = \frac{\hat{n}_i + s \sum_{r} \hat{n}_j } {\left| \hat{n}_i + s \sum_{r} \hat{n}_j \right|}
\end{equation}
The direction of the active force exerted on each CMC vertex $i$, $\hat{f}_i$, is a weighted average of the local outwards normal direction ($\hat{n}_i$) and a contribution from all the vertices $j$ within a distance $r$ from the vertex $i$ (and in the same connected cluster), with a weight of $s$.

\begin{figure}
    \centering
    \includegraphics[trim={0.5mm 0.5mm 0.5mm 0.5mm},clip]{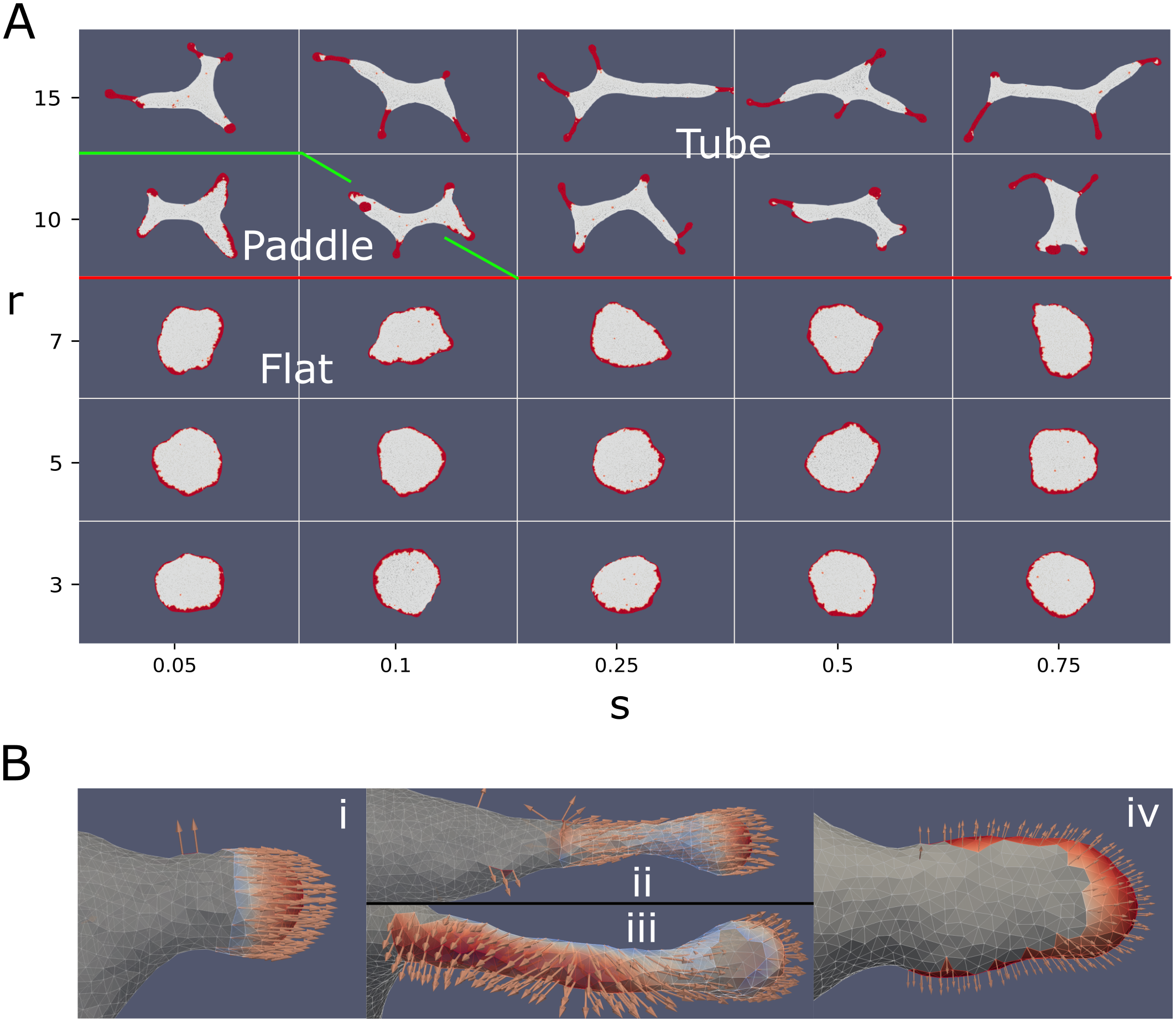}
    \caption{(A) Vesicle steady-state shapes as function of the strength ($s$) and range ($r$) of the Vicsek-like alignment interaction (Eq.\ref{eq:vicsek_force})($\rho=20\%$,$\kappa=28.5$,$C_0=0.4$,$w=2$,$f=0.2$). Interaction radius smaller than 10 leads to a flat phase. Above an interaction radius of $10$, the system transitions from a flat to a tubes phase. In between the flat and elongated tubes phases, we find a phase with "paddle"-like clusters. The tubular phase is characterized by CMC clusters that are mostly finger-like with a bulbous tip and a tubular sleeve, which often stretch a membrane tube behind them. (B) Snapshots of CMC clusters, with the active forces indicated by the arrows, and the colormap indicating the dot product of the local force and local outwards normal. In the tubes phase ($s=0.75$, $r=15$) we show in (i) an example of a hemispherical cluster, which is not able to pull an elongated protrusion. In (ii) (top) we show an example of a CMC cluster that contains a tubular sleeve, which increases the net pulling force above the threshold to pull a membrane tube. Note that at the sleeve base the alignment is weak due to the bare membrane boundary. This effect is also shown in (iii) (bottom), where a small patch of bare membrane is trapped between the cluster tip and the sleeve, leading to formation of two different alignment domains within the same cluster. Finally, in (iv) we show an example of the paddle cluster ($s=0.1,r=10$), where the weak alignment interaction gives rise to shapes similar to the regular arc-like clusters (Fig.\ref{fig:4_cf_and_coral}A), elongated by the non-normal force.}
    \label{fig:6_Aligned_force_phase_diagram}
\end{figure}

In Fig.\ref{fig:6_Aligned_force_phase_diagram}A we plot typical steady-state snapshots of the vesicle shape and CMC clusters, as function of the strength and range of the alignment interaction of Eq.\ref{eq:vicsek_force}. We observe a rather sharp transition from flat shapes for short-range alignment ($r<10$) to shapes containing thin tube-like protrusions for long-range alignment. As function of the parameter $s$ we find only weak dependence: at very small values of $s$ and $r=10$, we find that the weak alignment is sufficient to increase the net pulling force of the CMC clusters, such that they break the circular rim of the flat shape (Fig.\ref{fig:4_cf_and_coral}B(iii)). The resulting shape, with "paddle"-like protrusions, resembles the "arcs" phase we found in Fig.\ref{fig:4_cf_and_coral} between the flat and tubes phases. At higher values of $s$ this paddles phase changes to tubes, due to the stronger alignment leading to a larger net pulling force. 

At these larger interaction strength the vesicle produces thin, finger-like clusters with a small bulbous "head" and an elongated "sleeve" (Fig.\ref{fig:6_Aligned_force_phase_diagram}B(ii)). This shape allows the CMC to satisfy their spontaneous curvature, with a spherical tip that has a radius of $r_{tip}=2/C_0$, while the sleeve has a thinner radius of $r_{sleeve}=1/C_0$. Such a cluster configuration is stable due to the alignment of the active forces along the tube axis (perpendicular to the membrane along the sleeve, Fig.\ref{fig:6_Aligned_force_phase_diagram}B(ii)). Once these elongated clusters form, they exert a large pulling force on the remaining membrane, thereby pulling elongated bare-membrane tubes behind them. The membrane tube can have a larger radius than the radius of the tubular CMC cluster, as it balances the pulling force with the restoring force due to bending energy. The alignment of the forces means that the entire CMC cluster pulls along the protrusion axis (Fig.\ref{fig:6_Aligned_force_phase_diagram}B(ii)), exerting a much larger total force than was possible using purely normal forces at the tip, thereby forming tubes at values of the force per protein that are much lower than predicted by Eq.\ref{eq:force_balance} and Fig.\ref{fig:4_cf_and_coral}A. Smaller clusters that only contain the hemispherical tip (such as Fig.\ref{fig:6_Aligned_force_phase_diagram}B(i)), do not grow tube-like protrusions, even though their net pulling force is larger by up to a factor of $2$ compared to normal-force CMC, due to alignment (compare Fig.\ref{fig:6_Aligned_force_phase_diagram}B(i) to Fig.\ref{fig:4_cf_and_coral}B and Eq.\ref{eq:Fpull}). 

\begin{figure}
    \centering
    \includegraphics[trim={0.5mm 0.5mm 0.5mm 0.5mm},clip]{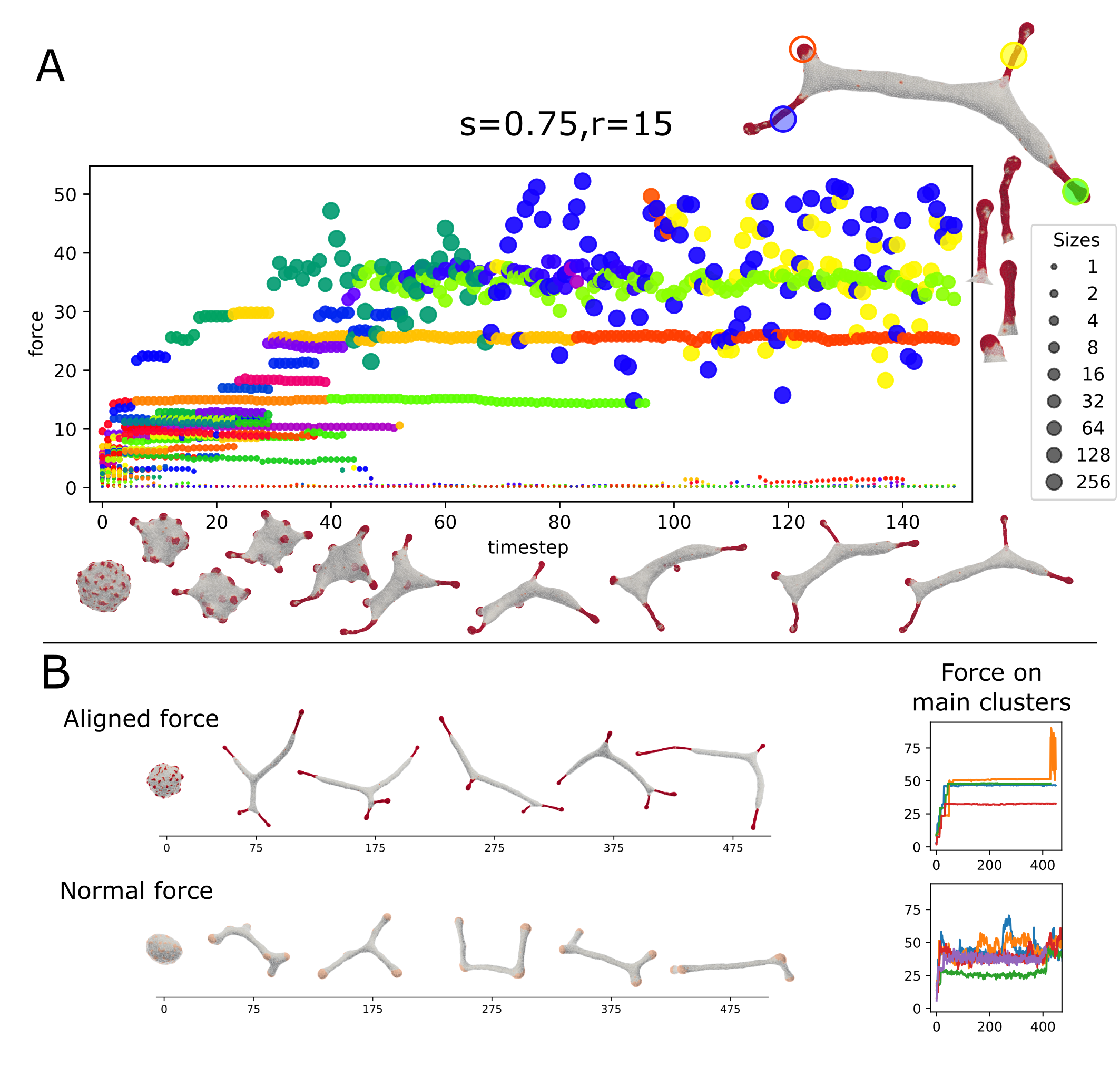}
    \caption{(A) Dynamics of the formation of the tubular phase, driven by strong alignment interactions ($\rho=20\%,\kappa=28.5,C_0=0.4,w=2,f=0.2,s=0.75, r=15$). Each circle represents a CMC-cluster at different MC time (x axis), the y axis represents the total force exerted by the cluster. The circle size represents the size of the CMC cluster (see sidebar). Color gives a persistent "identity" to each cluster, which last until fusion or fission. On the top right is a snapshot of the vesicle in the last time step. The four largest cluster are highlighted, and also shown on the right of the panel. Below the x-axis, we give snapshots of the vesicle. The rapid initial formation of buds is seen followed by slower fusion of clusters to form elongated protrusions. Two of the final large clusters, the bud and one of the elongated tube, are relatively stable, while the other two elongated clusters have wildly oscillating force. We can see on the right that the fluctuating cluster incorporates a few bare membrane vertices (Fig. \ref{fig:6_Aligned_force_phase_diagram}B,iii). (B) The dynamics of tube formation due to aligned force with highly curved CMCs (top, $s=0.5$, $r=30$, $f=0.2$, $C_0=0.4$) compared to formation due to shallow (weakly curved) CMCs with normal force (bottom, $f=0.5$, $C_0=0.1$). The tubes of the latter are more dynamic and less stable than clusters of the former. This is also seen on the right panel, which shows the total force on the largest clusters, which is far less noisy for the former.}
    \label{fig:7_Aligned_progression}
\end{figure}

In Fig.\ref{fig:7_Aligned_progression}A we plot the time progression of a vesicle with aligned-force CMC. We observe that initially localized hemispherical buds form rapidly. These buds then coalesce to form larger clusters that grow into the typical shape of bulbous tip with a thinner tubular part behind it. The size and total force of each of the clusters are plotted as function of time, with each point size indicating the cluster size, and its y-axis coordinate giving its total active force, respectively. Note that clusters that contain patches of "trapped" bare membrane undergo large force fluctuations (blue and yellow points, largest two clusters shown on the right of Fig.\ref{fig:7_Aligned_progression}A). These fluctuations arise from loss of global alignment over the entire CMC cluster, due to the bare membrane patch that allows the alignment to change, especially between the protrusion tip and the tubular part.

In Fig.\ref{fig:7_Aligned_progression}B we compare the finger-like protrusions that form due to highly curved aligned-force CMC, with the tubular shapes that form due to weakly curved normal-force CMC (Fig.\ref{fig:4_cf_and_coral}A). The main difference is that the aligned-force protrusions are much more stable compared to the tubes formed by the much smaller clusters of normal-force CMC. The normal-force CMC undergo frequent fission and coalescence events, that correspond to tubes shrinking and regrowing. These differences in dynamics can be seen in the SI movies S1,S2.

\section{Vesicles with both normal and aligned-force CMC, adhered to a flat substrate}

\begin{figure}
   \centering
   \includegraphics[trim={0.5mm 0.5mm 0.5mm 0.5mm},clip]{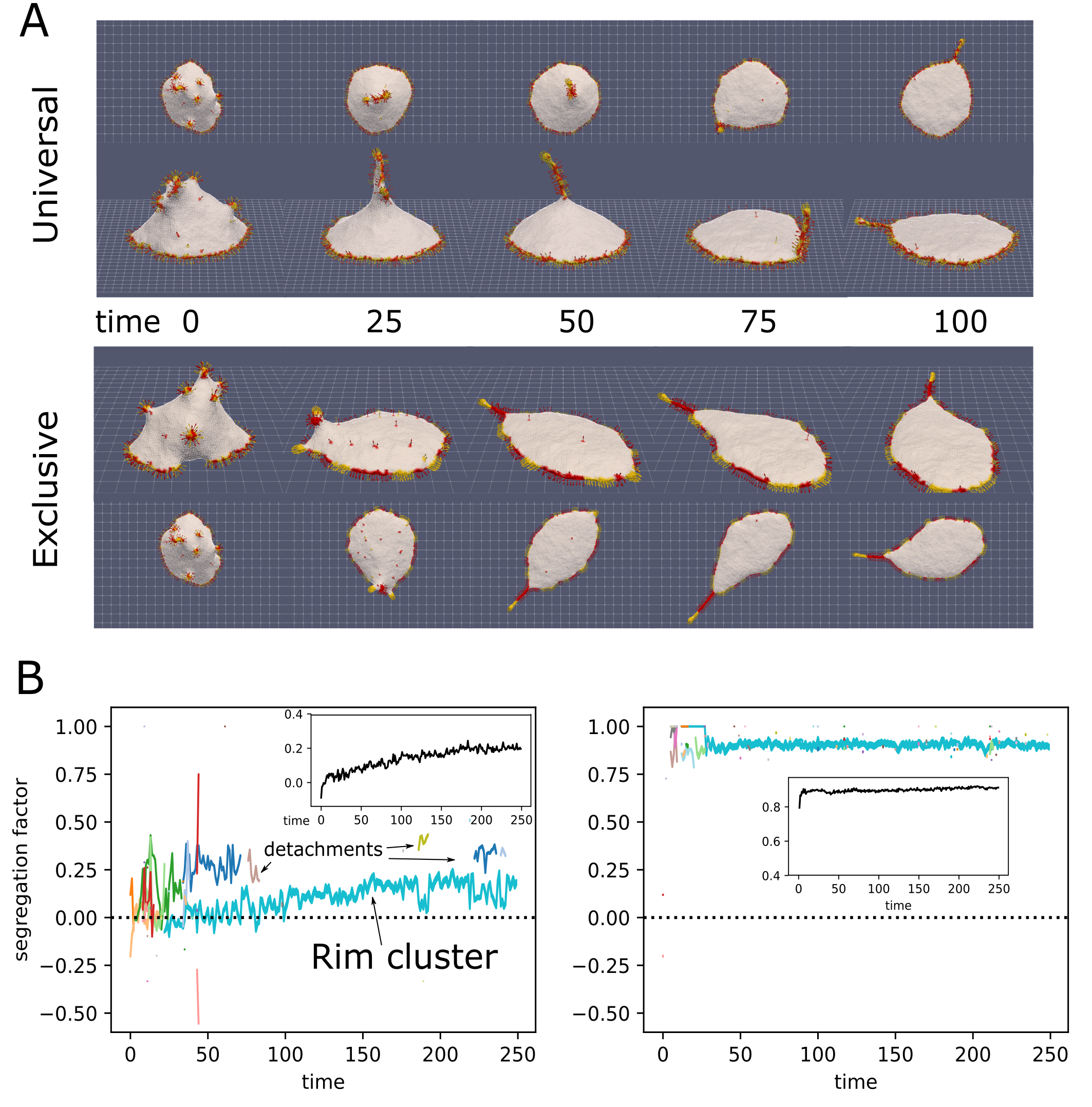}
   \caption{A: Initial progress of simulation with normal-force CMCs (red) and aligned-force CMCs (yellow), in universal binding (top) and type-exclusive binding (bottom), from the side and above ($\rho_{align}=10\%$, $\rho_{normal}=10\%$, $\kappa=28.5$, $C_0=0.8$, $w=2$, $f=0.5$, $s=0,1$, $r=15$, $w_{ad}=0.25$). CMCs in the rim drive the spreading of the vesicle on the surface, while some aligned-force CMCs aggregate into a bulb-and-sleeve cluster which drives the formation of a filopodia-like protrusion. This protrusion can attach to the rim cluster and then adhere to the substrate, while it can also detach from the substrate, and consequently also from the rim cluster. B: Evolution of the segregation factor in the simulations (Eq.\ref{segregation}). The colored lines give the segregation factor for each cluster, with the cluster size indicated by the line thickness. In the inset we give the average of the total segregation factor for 4 independent simulations. In the universal binding simulation we can see the fliopodia-like cluster repeatedly attach and detach from the rim cluster. The rim cluster is mostly mixed for this case, while the protrusion is much more segregated, as its tip is enriched with aligned-force CMCs.}
   \label{fig:8_Adhesion_with_alignment}
\end{figure}

We simulate a vesicle with a mixture of CMCs ($\rho=5\%$ of each type), both highly curved and convex, one type with normal force and the other with strongly aligned force ($r=15,s=1$). Our initial state of the vesicle is obtained by letting the vesicle spread over a flat adhesive substrate, while it contains only normal-force CMC. Then, at a time where the vesicle is partially spread (time $0$ in Fig.\ref{fig:8_Adhesion_with_alignment}A), we convert randomly half of the CMC to aligned-force behavior. We chose an adhesion strength $w_{ad}=0.25$ (Eq.\ref{eq:2}), which gives a well-spread vesicle when containing only normal-force CMC \cite{Sadhu2021}.

In Fig.\ref{fig:8_Adhesion_with_alignment}A we show two simulations: one with universal binding between the normal and aligned-force CMCs, and the other with exclusive binding, such that normal-normal and aligned-aligned CMC bind to their own type exclusively. In these examples we see that the rim cluster forms and drives strong spreading of the vesicle, as expected \cite{Sadhu2021}. The aligned-force CMC (labeled in yellow) aggregate to form a single filopodia-like protrusion, which is able to recruit into it also normal-force CMC (labeled in red). This filopodia is highly dynamic, undergoing periods of attachment to the rim cluster, and to the adhesive substrate, as well as detachments from the substrate. The filopodia is observed to attach and detach from the rim cluster, leading to meandering motion. When the two types bind exclusively, they form segregated clusters along the rim, with the aligned-force clusters protruding slightly more outwards compared to the normal-force clusters. The dynamics of this system can be seen in SI movie S3.

In Fig.\ref{fig:8_Adhesion_with_alignment}B we show the evolution of the segregation factor in the simulations, which is defined as 
\begin{equation} 
S=2 \cdot\text{Prob}\left( \text{CMC neighbor is of the same type} \right)-1
\label{segregation}
\end{equation}
This segregation factor is equal to $0$ for well-mixed clusters (where the probability to have a neighbor CMC of the same type is equal to $1/2$), and it is equal to $1$ for complete phase-separation of the types. In the main panels we give the segregation factor per cluster for the simulations shown in Fig.\ref{fig:8_Adhesion_with_alignment}A. The insets show the average of $4$ independent simulations, which converge to a value of about $S=0.25$ for the universal binding and $S=0.9$ for the exclusive binding. In the universal case, we can see that the segregation is strongest in the filopodia, so the segergation factor for the large rim cluster jumps up or down, when the filopodia protrusion cluster attaches or detaches respectively. The protrusion cluster is more segregated ($S\approx0.25$), since its tip is enriched with aligned-force CMCs that drive its formation, while the rim cluster is nearly perfectly mixed ($S\approx 0$). For the exclusive binding, the segregation is high both in the filopodia protrusion and in the rim cluster, so it does not change when the filopodia attach or detach from the rim.

Note that along the adhered vesicle rim, the regions of aligned-force CMC protrude slightly more than the normal-force regions (Fig.\ref{fig:8_Adhesion_with_alignment}A, exclusive). This is enhanced when the normal-force CMC are disabled, so that they do not exert any active force, as shown in Fig.S6. 

\section{Comparison with Experiments}

We can now compare some of our theoretical results to experimental observations, published and new.

\subsection{Membrane shapes driven by branched actin polymerization}

The active protrusive forces in our model are representative of actin polymerization activity near the cell membrane. When the actin polymerization is nucleated by proteins that induce branched actin networks (such as WASP, WAVE \cite{small2002lamellipodium,takenawa2007wasp,rottner2019assembling}), it is more natural to describe the force as a local pressure on the membrane, which therefore acts towards the outwards normal. 

The variety of shapes we obtained in our model (Figs.\ref{fig:2_FW_Phase_Diagram},\ref{fig:3_FW_progression}), range from flat lamellipodia-like shapes, to cylindrical filopodia, to pearling-like protrusions. Some of these new elongated shapes can be compared with elongated global cell shapes,  observed in living cells \cite{suetsugu2006rac}.

\subsection{Membrane shapes driven by bundled actin polymerization}

The introduction of alignment in the forces exerted by the CMC represents in our model the case of proteins that nucleate parallel actin bundles, such as VASP and Formins \cite{Faix2006,vaggi2011eps8,Feng2022}. Our model has demonstrated previously that curved proteins that apply normal forces, induce the formation of flattened, lamellipodia-like protrusions \cite{Fosnaric2019,Sadhu2021}. Here we show that curved proteins that induce polymerization of bundled actin (aligned-force in our model), naturally give rise to filopodia-like protrusions (Figs.\ref{fig:6_Aligned_force_phase_diagram},\ref{fig:7_Aligned_progression}). This result fits the observation of highly curved convex-shaped proteins such as IRSp53 in both the leading edge of lamellipodia \cite{pipathsouk2021wave,galicReview2022} and in filopodia \cite{Mattila2008}, where the actin organization is very different due to the different type of actin nucleators \cite{small2002lamellipodium,krause2014steering}. Note that the combination of convex curvature, and nucleators of bundled actin, can form filopodia even without the explicit presence of I-BAR proteins (such as IRSp53) \cite{kuhn2015structure,pokrant2023ena}.

Note that protrusions of similar shapes to our aligned-force protrusions, which have a bulbous tip and a slender neck (Figs.\ref{fig:6_Aligned_force_phase_diagram},\ref{fig:7_Aligned_progression}), were theoretically predicted to form by anisotropic CMC, even without force \cite{bobrovska2013role}. Similar thin tubes with bulbous tips are observed in cellular nanotubes \cite{schara2009mechanisms} and in filopodia \cite{miihkinen2021myosin}. Since many curved proteins, such as IRSp53 are anisotropic in their intrinsic shape, it will be interesting to extend our work in the future to include such anisotropy.

Finally, our simulations of an adhered vesicle (Fig.\ref{fig:8_Adhesion_with_alignment}) indicate that the filopodia protrusions can undergo attachment and detachment from the substrate, resembling such motion observed in experiments \cite{lee2023lamellipodia}. In addition, when we mixed the aligned-force and normal-force CMC with exclusive binding between them, we obtained their segregated organization along the rim of the adhered vesicle. This is reminiscent of the observations of segregated regions of bundled actin and branched actin nucleators along the rim of cellular protrusions extending on adhered substrates \cite{cohan2001role,damiano2020loss,kage2022lamellipodia,faix2022ena}. As in the experiments, the clusters of aligned-force CMC along the rim slightly protrude, as they exert a higher local force on the membrane rim, compared to the normal-force CMC. These small protrusions have been termed "spikes" and "microspikes" along the edge of lamellipodia in cells \cite{koestler2008differentially,damiano2020loss,pokrant2023ena}.

\begin{figure}
   \centering
   \includegraphics[trim={0.5mm 0.5mm 0.5mm 0.5mm},clip]{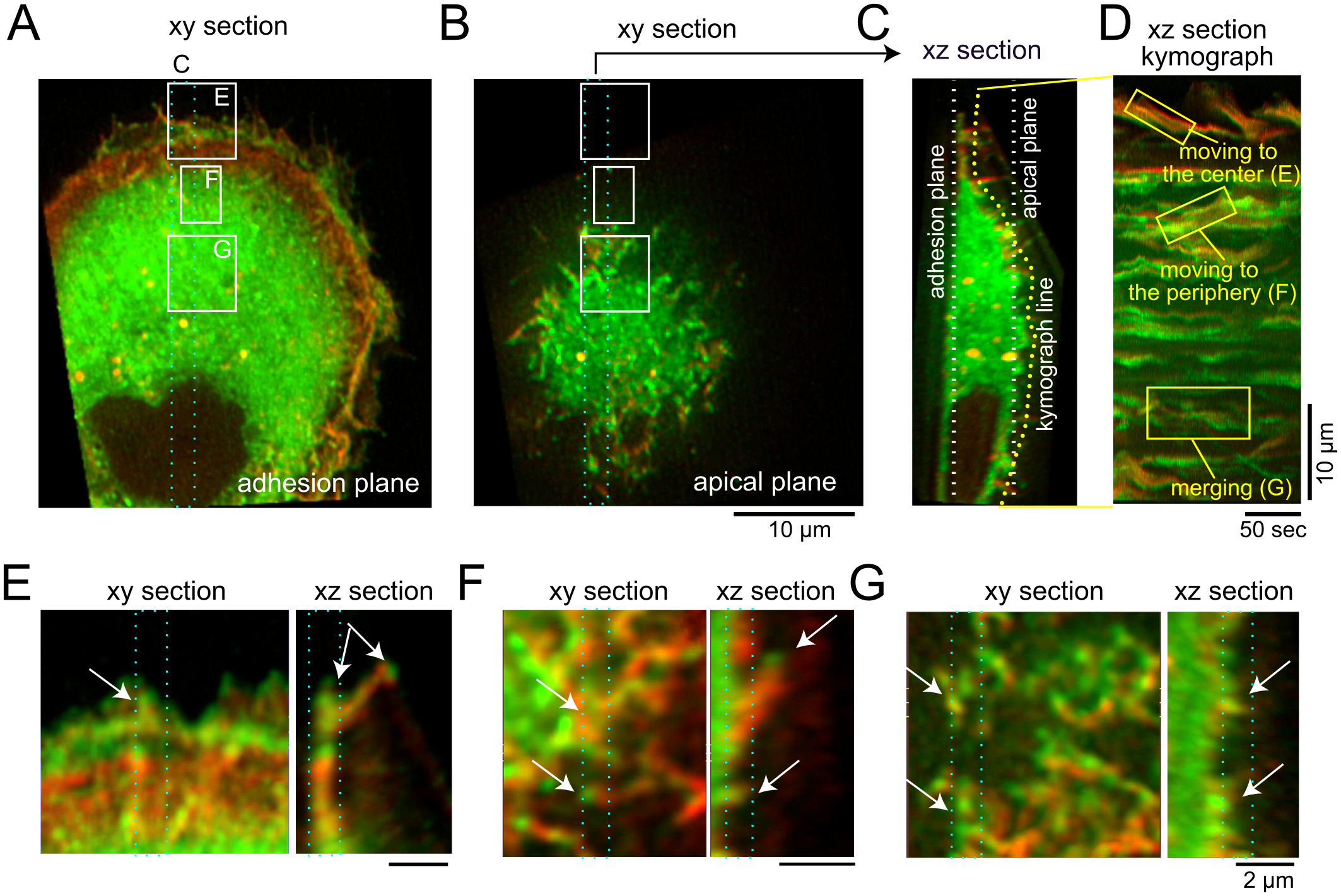}
   \caption{Movements of IRSp53-localized cellular protrusions. (A, B) The adhesion (A) and apical (B) plane section of the three-dimensional images of an IRSp53-knockout U251 glioblastoma cell expressing GFP-IRSp53 (green) and mCherry-lifeact (red). In (A) and (B), the region for the $\sim 2 $ µm thick xz section projection is indicated by the cyan dotted rectangle. (C) The xz section of the region of (A). The white lines indicate the plane in (A,B). The yellow line, which was set in the proximity of the surface plane of the cell, indicates the line for the kymograph. (D) The kymograph of the cell surface as indicated in the yellow line in (C), along with the annotation of the representative motion of the IRSp53. (E-F) The xy and xz sections at the regions that are marked in (A,B), from the periphery (E), the middle (F), and the center (G). The plane parallel to the plasma membrane was sectioned and the regions that were projected xy and xz sections each others were marked in cyan dotted rectangles. Arrows indicate the protrusions. The scale bar, $10$ µm (A-D), $2$ µm (E-G), and $50$ sec (D).}
\label{fig:9_Experiment}
\end{figure}

In Fig.\ref{fig:9_Experiment} we show images illustrating the dynamics of filopodia in cells, using lattice light-sheet microscopy, which is capable of the high spatial and temporal resolution necessary to view the dynamics of the thin filopodia \cite{mimori2023imaging}. The curved membrane protein IRSp53 is fluorescently labeled in green (GFP-IRSp53), while the actin filaments are labeled in red (mCherry-lifeact). We observe in the experiments several features that are captured by the theoretical model: The filopodia are highly dynamic, both at the cell rim and along its dorsal surface (Fig.\ref{fig:9_Experiment}A-D), as we also see in the model (Fig.\ref{fig:8_Adhesion_with_alignment}). The filopodia in the experiments migrate on the cell surface, merge with other filopodia, and undergo attachments and detachments from the surface (see SI movies 5-8), as we also see in the simulations (SI movies 3 and 4). Our assumption in the model of uniform adhesion along the membrane, and along the filopodia, agrees with some observations \cite{miihkinen2021myosin,tu2022filopodial}, and we can add more complex adhesion models in the future if needed. Note that in the cells we observe an additional retraction motion that is driven by myosin-II contractile forces, which we do not have in our current model. 

The highly curved IRSp53 is observed to aggregate strongly at the tips of the filopodia, while along the lower parts of the protrusion its aggregation is more fragmented (Fig.\ref{fig:9_Experiment}E,F). This fits with the shapes that we obtained in the model (Fig.\ref{fig:6_Aligned_force_phase_diagram}B,\ref{fig:8_Adhesion_with_alignment}A). Furthermore, our simulations of mixtures of aligned-force and normal-force CMC indicate that while the aligned-force CMC are essential for forming the filopodia protrusions and occupy its tip region, there can be significant amount of normal-force CMC along the lower part of the filpodia. Since the normal-force CMC correspond to branched-actin nucleators, this result suggests that along the lower part of filopodia we may expect to find proteins such as WAVE, which are usually associated with the leading edge of the lamellipodia. This prediction is supported by some experimental observations of WAVE proteins \cite{nozumi2003differential}, Arp2/3 complexes \cite{johnston2008arp2}, and small lamellipodia-like protrusions, along filopodia shafts \cite{lebrand2004critical}.

\subsection{Membrane shapes driven by mixtures of passive concave and active convex CMC}

Our mixtures of CMC of opposite curvatures (Figs.\ref{fig:4_cf_and_coral}C,\ref{fig:5_Coral_sack_shapes}) gives rise to membrane shapes that resemble in their texture the ruffles observed in cells \cite{Sitarska2021}. In addition, we find that when the passive concave component is highly curved, we observe a phase separation within the CMC clusters, whereby the concave CMC forms an internalized spherical invagination. These invaginations are then squeezed at their base by the active forces induced by the convex CMC, and the calculated membrane shape dynamics resembles the process of actin-dependent endycytosis \cite{doherty2009mechanisms,mooren2012roles,moreno2021cargo,kaplan2022load} and macropinocytosis \cite{kay2021macropinocytosis,sonder2021restructuring,mylvaganam2021cytoskeleton,lutton2022formation}. 

Note that there is some experimental evidence that the internalized membrane, corresponding to our concave CMC region, do indeed contain concave membrane components, such as BAR proteins \cite{baranov2019chasing}. In addition, there are examples where the internalized region contains membrane components that interact with the convex proteins that recruit actin and form the squeezing at the narrow neck. In \cite{moreno2021cargo} the internalized activated integrins and associated proteins, bind to the actin which is nucleated at the neck, recruited there by IRSp53 (convex) proteins. In our model we show that such a direct interaction is necessary for robust formation of the internalized sacks with the recruited convex proteins at the neck.

\section{Discussion}

In this study we greatly extend our theoretical understanding of the space of membrane shapes that are produced by curved membrane protein complexes (CMC) that exert active protrusive forces on the membrane \cite{drab2023modeling}. We started by mapping the phases as function of the magnitude of the active force and attractive nearest-neighbor interaction strength of CMCs (Fig.\ref{fig:2_FW_Phase_Diagram}A), demonstrating the competition between these two terms: systems dominated by the binding interactions tend to have the equilibrium (pearled) shapes of the CMC clusters. The active forces tend to break-up the pearled clusters, and induce the formation of either elongated or flat pancake-like membrane shapes. 

Similarly we exposed the phase diagram in terms of the active force and the CMC spontaneous curvature (Fig.\ref{fig:4_cf_and_coral}A), whereby highly curved CMC induce flattened vesicle shapes, while less curved CMC induce elongated tubular shapes. Note that in these studies the protrusive force applied by each CMC is towards the local outwards normal. 

Based on these results we further explored systems where highly curved active CMC could induce tubular protrusions. We tested two possible scenarios: In the first one, the effective curvature of the CMC cluster is reduced by mixing two types of CMC of opposite curvatures, such that a tubular protrusion forms with a rather flat CMC cluster at its tip (Fig.\ref{fig:4_cf_and_coral}C,D). In the second, the net protrusive force of the CMC cluster is increased by introducing an alignment interaction that tends to align the forces exerted by CMC that are bound within the same cluster (Fig.\ref{fig:6_Aligned_force_phase_diagram}). This alignment is found to stabilize long tubular CMC clusters, since the aligned active forces act along the tube axis and do not act to expand the tube, unlike the case of normal protrusive forces.

We found that that mixtures of CMC of opposite curvatures, specifically passive concave and active convex, lead to formation of clusters with complex textures that resemble ruffles on cell membranes (Figs.\ref{fig:4_cf_and_coral}C,\ref{fig:5_Coral_sack_shapes}). In addition, we found in these systems the formation of internalized invaginations, where the convex active CMC form a narrow neck, resembling endocytosis and macropinocytosis in cells.

To conclude, the results presented in this work expand out theoretical understanding of membrane shapes and dynamics driven by intrinsic (spontaneous) curvature of membrane components and cytoskeletal active forces. Some of these shapes resemble observed membrane dynamics in living cells, suggesting that this coupling between curved membrane proteins and cytoskeleton forces gives rise to these biological phenomena. Many of the features that we found, such as the ruffles and the internalized invaginations by mixing CMC of different curvatures, remain to be further explored in future theoretical studies. In addition, future studies will explore the dynamics of the membranes when the CMC have anisotropic spontaneous curvature, and also in the presence of contractile forces.






\section*{Conflict of Interest Statement}

The authors declare that the research was conducted in the absence of any commercial or financial relationships that could be construed as a potential conflict of interest.

\section*{Author Contributions}
YR and NG developed the theoretical model; SP and AI developed the software; YR and NG conceived, designed and implemented the analysis of the model, and prepared the manuscript. YK and SS cultured and imaged the cells. The manuscript was edited by all the authors. 

\section*{Funding}
NG is the incumbent of the Lee and William Abramowitz Professorial Chair of Biophysics, and acknowledges support by the Ben May Center for Theory and Computation, and the Israel Science Foundation (Grant No. 207/22). AI and SM were supported by the Slovenian Research Agency (ARRS) through the Grants No. J3-3066 and J2-4447 and Programme No. P2-0232. 
YK and SS was supported by grants from the JSPS (KAKENHI JP20H03252, JP20KK0341, and  JP21H05047) and JST CREST (JPMJCR1863) to SS and  Takeda Science Foundation, a Grant-in-Aid for Challenging Exploratory Research (KAKENHI No. 20K20379), and  JST CREST (JPMJCR1863) to YK.

\section*{Acknowledgments}
NG is the incumbent of the Lee and William Abramowitz Professorial Chair of Biophysics. This research is made possible in part by the historic generosity of the Harold Perlman Family.

\section*{Supplemental Data}
 The SI text, figures, and movies are also available from the \href{https://weizmann.app.box.com/s/5z6aexxk1lmne0b29sf25o451unnj85x}{Box drive}.

\section*{Data Availability Statement}\label{section:data availability}
The code for generating the simulations of this study can be found in the \href{https://github.com/yoavrv/cluster-trisurf}{GitHub repository} of YR, which is taken and modified off the \href{https://git.penic.eu/summary/trisurf-ng.git}{GitBlit repository} of SP. Reconstruction of the initial simulation folders are also available from the \href{https://weizmann.app.box.com/s/5z6aexxk1lmne0b29sf25o451unnj85x}{Box drive}. Further data or code requests will be happily fulfilled by YR.

\bibliography{references}

\section*{}

\begin{table}
    \centering
    \begin{tabular}{|c|c|c|c|c|c|c|c|c|}
        \hline
         parameter & units & Fig.\ref{fig:1_Fosnaric_phases},\cite{Fosnaric2019} & Fig.\ref{fig:2_FW_Phase_Diagram} & Fig.\ref{fig:4_cf_and_coral}A & Fig.\ref{fig:4_cf_and_coral}C & Fig.\ref{fig:6_Aligned_force_phase_diagram} &
         Fig.\ref{fig:7_Aligned_progression} &
         Fig.\ref{fig:8_Adhesion_with_alignment}\\
         \hline
         $f$ & $K_B T / \ell_{min}$ & $1$ & $0-1.2$ & $0-0.5$ & $0.5$& $0.2$ & $0.2$, $0.5$ & 0.5 \\
         $w$ & $K_B T$ & $1$ & $0-4.8$ & $2$ & $2$ & $2$& $2$ & $2$\\
         $\kappa$ & $K_B T$ &  $20$* & $20$ & $28.5$ & $28.5$ & $28.5$ & $28.5$ & $28.5$ \\
         $\rho$ & 1 & $0\%$-$20\%$ & $10\%$ & $20\%$ & $10\%,10\%$ & $20\%$& $20\%$,& $10\%,10\%$\\
         $C_0$ & $1/\ell_{min}$ & $1$($0$) & $1$ & $0.8$ & -$0.75-0, 0.8$ & $0.4$ & $0.4$, $0.1$ & $0.8$\\
         \hline
    \end{tabular}
    \caption{The values of the model parameters used in the simulations, in the different figures. The energy units are $K_B T=1$, which define the scale of $f,w,\kappa$, and the length units are $\ell_\text{min}=1$, which define the scale of the vertex lattice, the force, and spontaneous curvature.}
    \label{tab:simulation_parameters}
\end{table}

\end{document}


\title {Supplementary Material}

\renewcommand{\thefigure}{S-\arabic{figure}}
\renewcommand{\thesection}{S-\arabic{section}}
\renewcommand{\theequation}{S-\arabic{equation}}

\maketitle

\section{Calculation of the perimeter of CMC clusters}

\begin{figure}
    \centering
    \includegraphics[width=\linewidth]{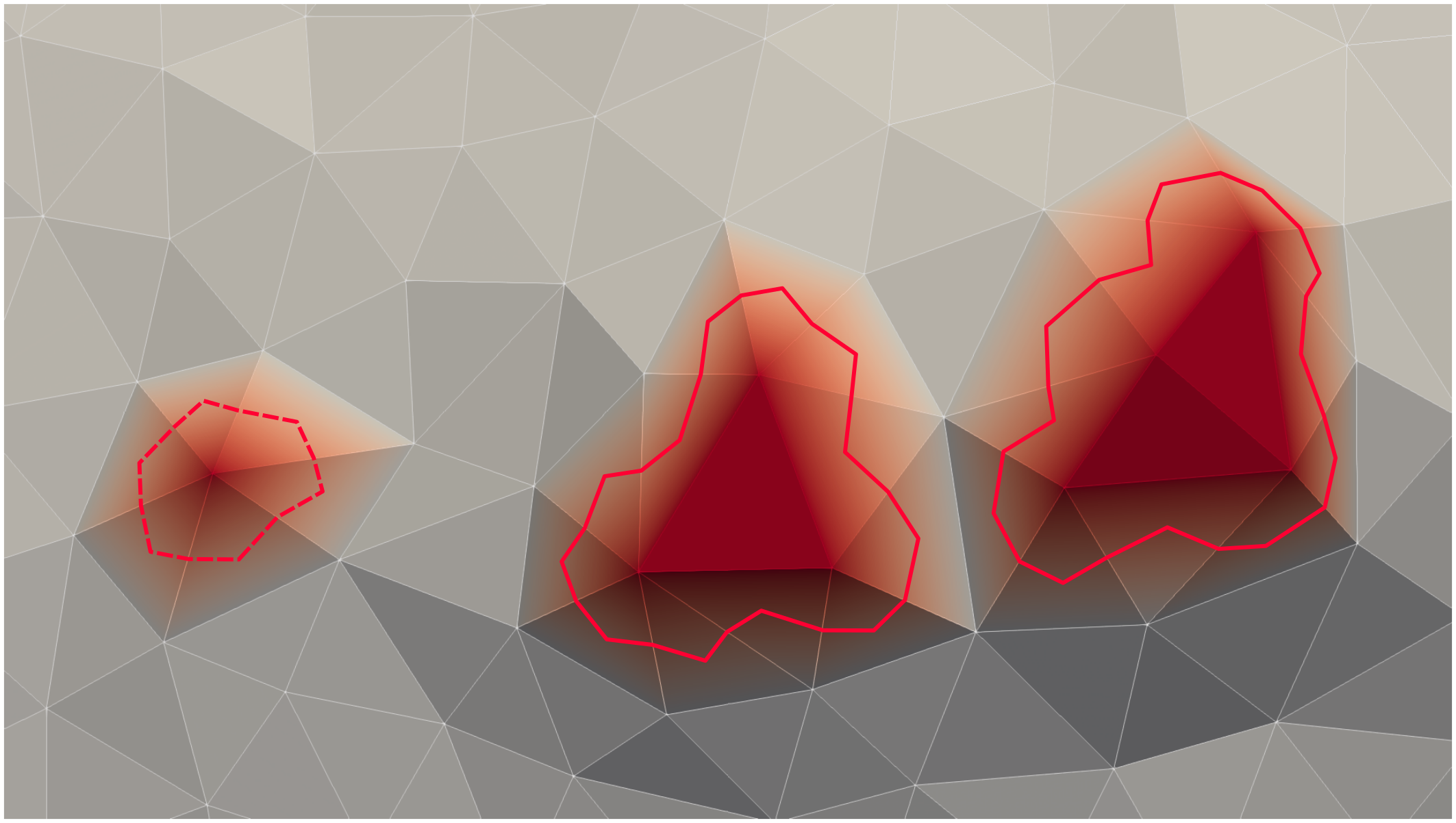}
    \caption{Sketch of the boundary of connected clusters: for each edge between the cluster and the outside, a line is drawn from the middle to the center each of the adjacent triangles. We ignore the single-clusters (dashed line)}
    \label{fig:1_perimeter_calculation}
\end{figure}

The CMC-bare membrane boundary is measured by summing the dual of the edges between the cluster and bare membrane. These are the edges in the voronoi lattice, connecting the mid-section of each edge to the circumcenter of the adjacent triangles i.e. the center of the inscribing circle (see figure \ref{fig:1_perimeter_calculation}). Partitioning each triangle between its vertices is already used in the calculation of the curvature \cite{Gompper2004}.

\section{Analytical calculation of the flat-pearling phase transition line}
\begin{figure}
    \centering
    \includegraphics[width=\linewidth]{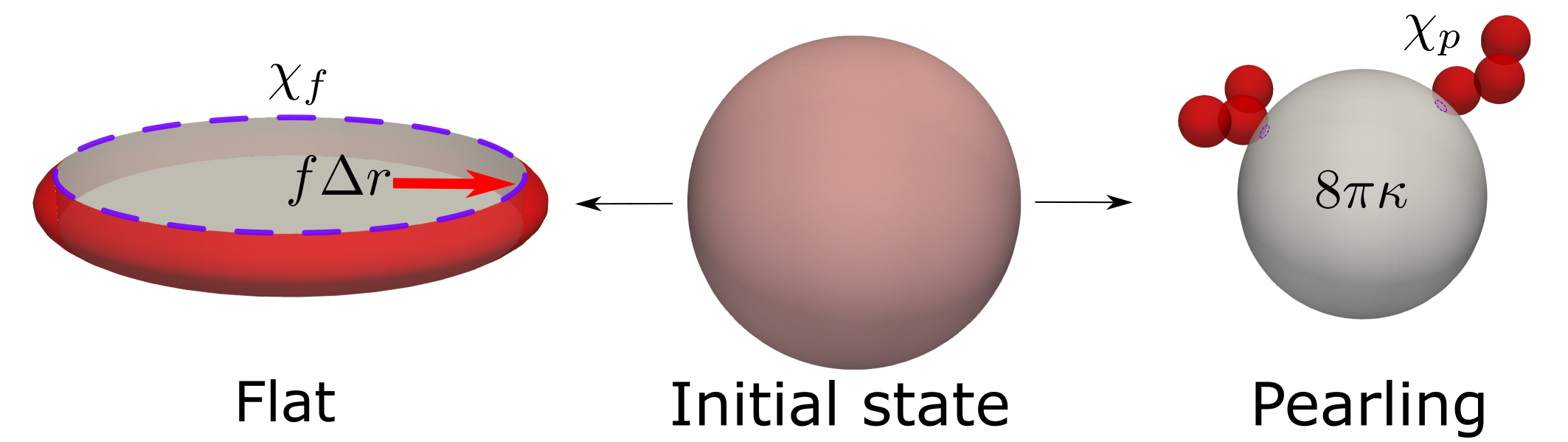}
    \caption{Schematic description of the transition between flat and pearling phases, from an initially mixed, spherical phase (at the center). Bare membrane is in white, and CMCs in red, and mixed composition in pink. The flat transition result in all CMCs moving from the surface of the sphere to the rim of a flat disc, which has a larger radius $\Delta r$. Due to active force $f$, this generates work $W=-f\Delta r$. The bending energy of the CMCs on the rim and in the pearling clusters is assumed to be approximately $0$, but the spherical body of bare membrane in the pearling phase has a bending energy of a closed sphere: $8\pi\kappa$, while it is zero for the flat discs of bare membrane in the flat phase (since they are flat). Finally, the number of CMC-CMC bonds in the pearling phase $\chi_p$ is larger than in the flat phase $\chi_f$, since in the flat phase it is reduced due to the large boundary between the rim cluster the the flat bare membrane discs.}
    \label{fig:2_flat_vs_pearling}
\end{figure}

We can make a rough analytical estimation for the flat-pearling transition by equating the active work and energy of the flat phase from a mixed phase to the energy of the pearling phase (Fig. \ref{fig:2_flat_vs_pearling}). In the flat phase, moving the active CMCs outwards from the radius of the sphere $r_p$ to the larger radius of the flattened disc $r_f$ results in work. The pearling phase has binding advantage because all CMC vertices are connected, with $-w$ per edge, while the flat rim has large interface (boundary perimeter length) where CMCs vertices neighbor bare membrane vertices, whose edge does not contribute. The pearling phase has a bending disadvantage due to the bare membrane body, which is roughly spherical with an energy of $8\pi\kappa$, compare to the flat phase where the bare membrane is in two flat discs with no bending energy (both the pearling and rim clusters are curved to fit the CMCs, so they do not have bending energy).

\begin{equation}
-\left(r_f - r_p \right) F =  -w\left(\chi_p - \chi_f\right) + 8 \pi \kappa 
\label{eq:greenLine}
\end{equation}

\begin{figure}
    \centering
    \includegraphics[width=\linewidth]{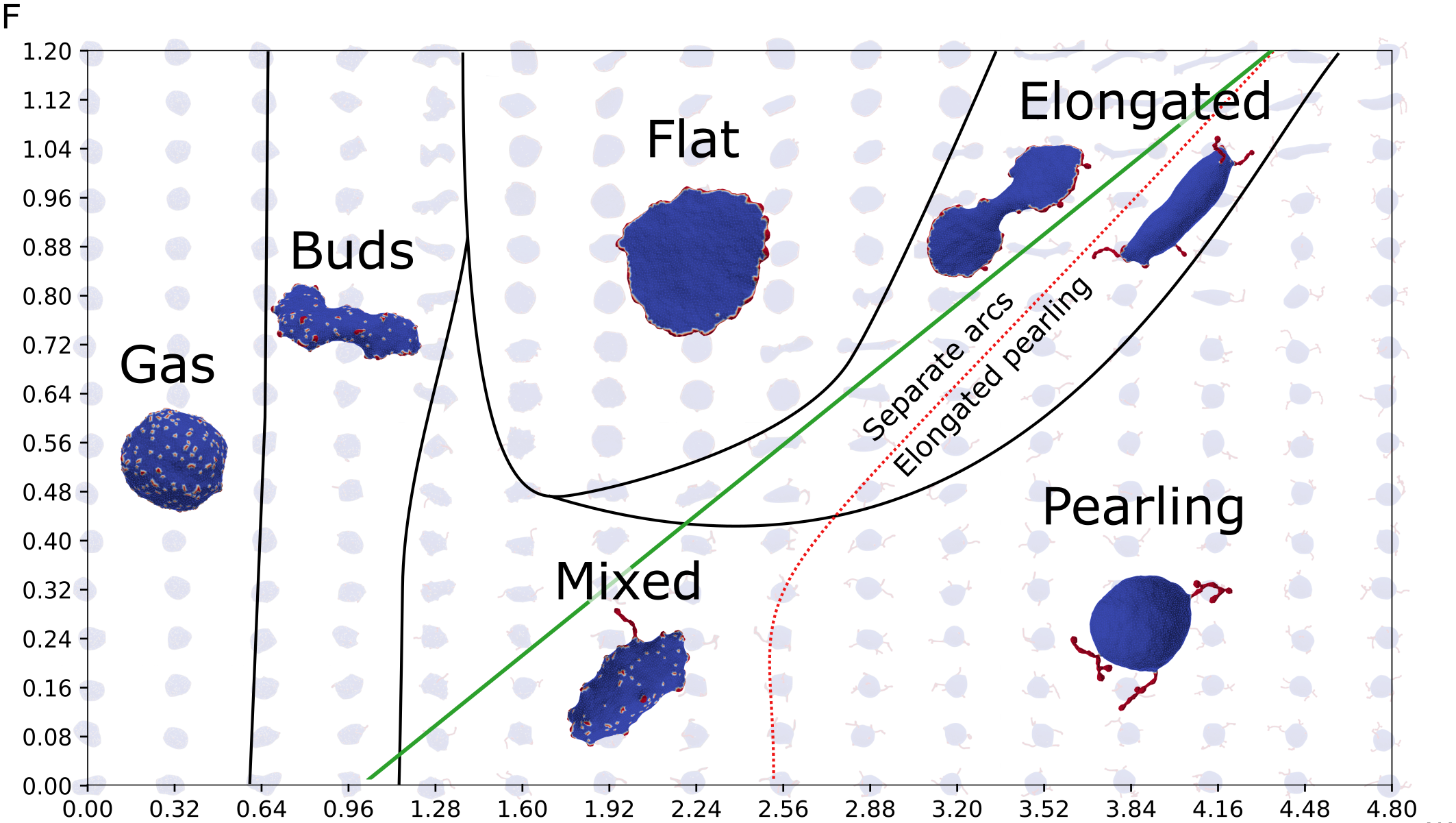}
    \caption{Phase diagram of the force-binding strength system, with an analytically-derived transition line for the pearling-flat transition (green line, Eq.\ref{eq:greenLine}). }
    \label{fig:3_analytical_line}
\end{figure}

The radius difference $\Delta r=r_f - r_p$ (Fig. \ref{fig:2_flat_vs_pearling}), and the number of CMC-CMC bonds $\chi_p,\chi_f$ in the pearling and flat phases respectively, are dependant on the geometry of the phases, so they should be very weakly dependant on the specific model parameters. Therefore $\Delta r$ and $\chi_p - \chi_f$  do not depend on $w,f,\kappa$, and we end up having a linear relation between $f$ and $w$ along the transition line in the $f,w$ phase diagram. In the force-binding strength ($f-w$) system, we take the values for these geometric quantities from simulations and draw the resulting line on the phase diagram (Fig. \ref{fig:3_analytical_line}, green line), which qualitatively matches the behavior of the transition observed in the simulations.


\section{Mixed curvature CMC clusters}

\begin{figure}
    \centering
    \includegraphics[width=\linewidth]{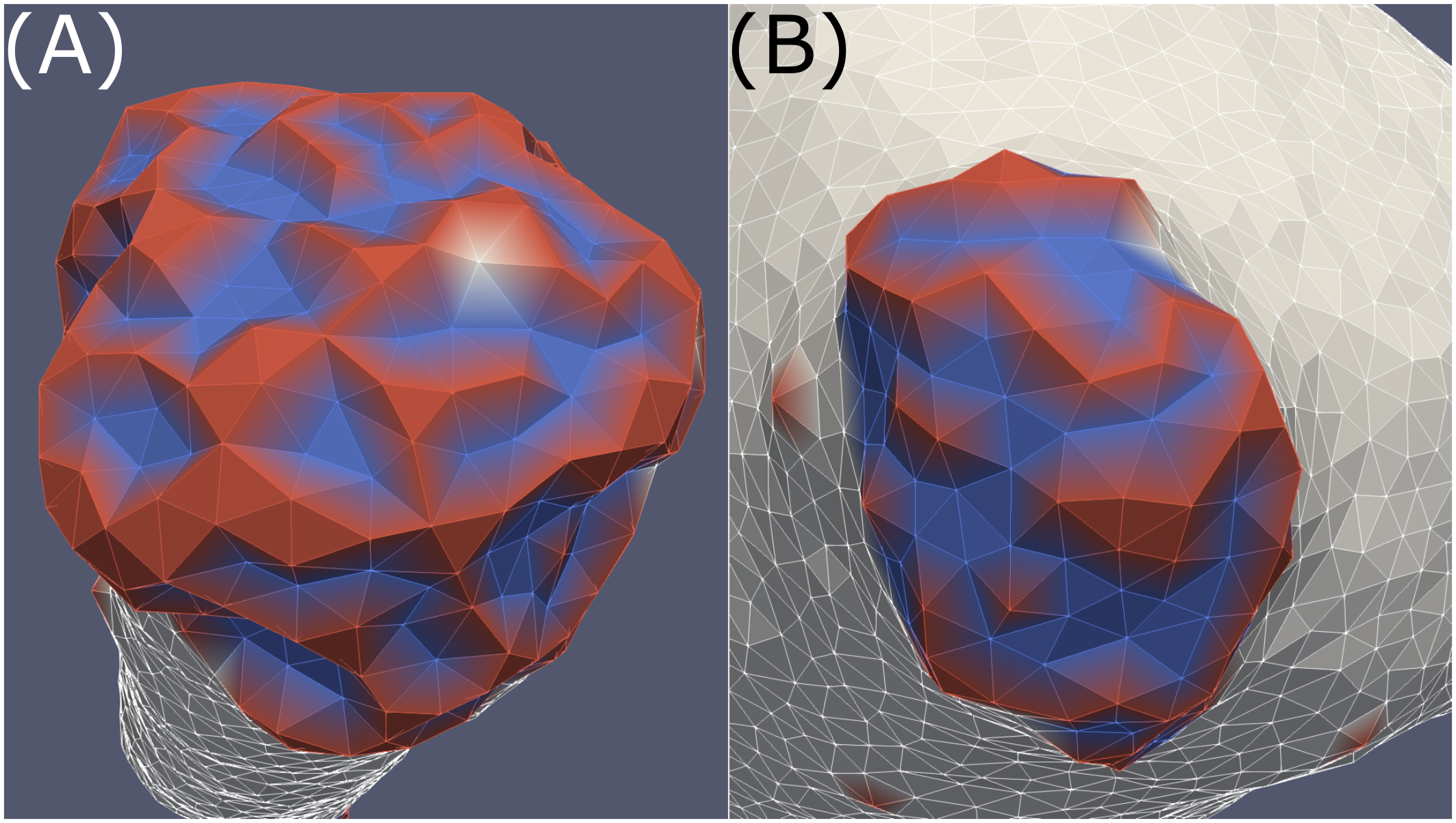}
    \caption{The undulation of a CMC cluster with (A) highly concave ($-0.6$) active CMC (B) with shallow concave ($-0.001$) CMC and disabled force. The size and shape of the clusters is very different, but the peaks and troughs patterning due to CMC shape is at the limit of the mesh resolution for both.}
    \label{fig:4_undulation_comparison}
\end{figure}

The concave and convex CMCs generate a wavelike pattern, but analyzing it in terms of wavenumber is difficult, since the clusters are part of an irregular, triangulated surface. The undulations of the CMCs in the mixed clusters are essentially independent of $C_0$, and $f$, as shown in Fig.\ref{fig:4_undulation_comparison}. Note that we are at the limit of the mesh resolution for these undulations. We have yet to be able to compare this to the experimental results in \cite{Sitarska2021}.

\section{Mixed Curvature With Exclusive binding}
\begin{figure}[ht]
    \centering
    \includegraphics[width=\linewidth]{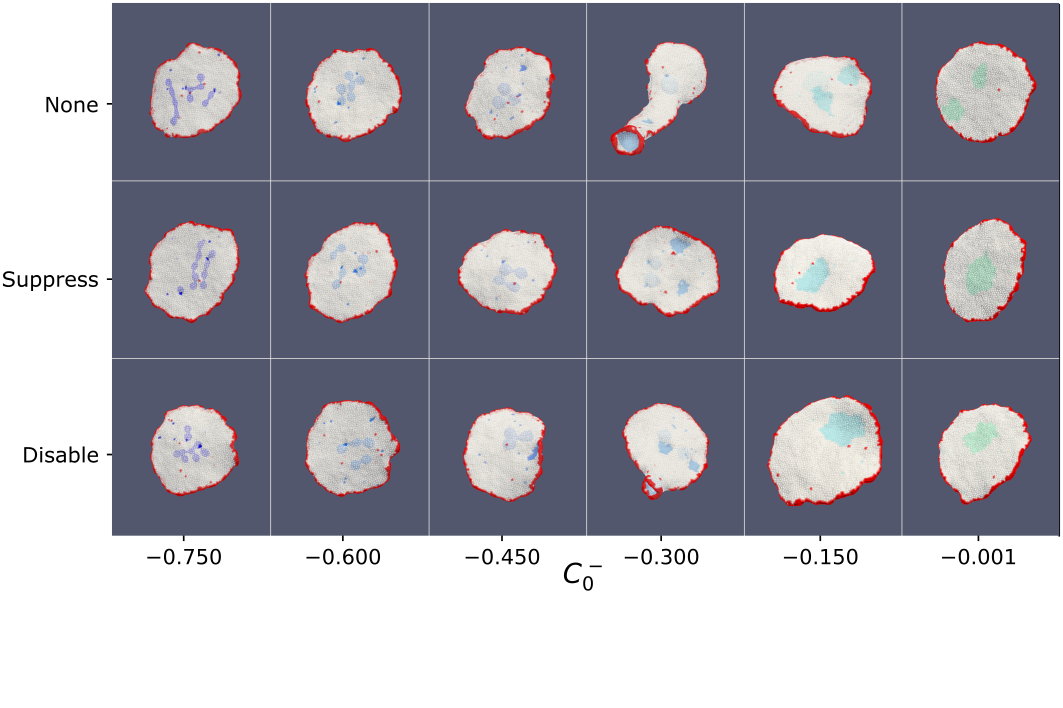}
    \caption{Active convex and passive concave system (red and blue, respectively), with binding between same type only. As in the universal binding case, the suppressive and disabling inhibition do not have any strong effects, since the types are separated. Simulations with $C_0^-\le-0.3$ are draw semi-transparent. In all cases, the convex CMCs aggregate in a rim, making the vesicle flat, and concave CMCs aggregate in pearling for $C_0^-<-0.3$, bowl-like patches for $C_0^->-0.3$, and both for $C_0^-=-0.3$. }
    \label{fig:5_Coral_exclusive_phase_diagram}
\end{figure}

The mixed curvature system (Fig. 4c in the main text) was also simulated using exclusive binding, i.e. only same-curvature CMCs bind together (Fig.\ref{fig:5_Coral_exclusive_phase_diagram}). The result is that the two CMCs types form separated aggregates, with the active convex CMCs aggregating along the rim and forming the flat phase. The passive concave CMC form separated clusters of different shapes, depending on their spontaneous curvature. Highly concave CMCs ($C_0^-\le-0.45$) aggregate into internal pearling clusters, that do not affect the flat global phase. The shallower concave CMCs ($C_0^-\ge-0.3$) aggregate into large, shallow bowl-like patches. 

In some cases, these concave aggregates are able to form with convex CMC along their rim, since their curvatures complement each other (see for example at $C_0^-=-0.3$). Since the convex active CMC along the rim of the concave cluster apply protrusive forces, they end up forming together a "cup"-like protrusion. When the force is inhibited, this aggregation occurs, but it is not elongated as a protrusion (compare "None" with "Disable" at $C_0^-=-0.3$ in Fig.\ref{fig:5_Coral_exclusive_phase_diagram}). Other than that, inhibition doesn't appear to significantly affect the results in Fig.\ref{fig:5_Coral_exclusive_phase_diagram}, since there is no significant contact between the two CMC types. These shapes, in the form of open bowls, resemble early stages of macropinocytosis \cite{veltman2016plasma,kay2021macropinocytosis}, but do not evolve to induce closure of the "mouth", as we observed when the convex and concave CMC had direct interactions (Fig.5 in the main text).

\section{Vesicles with both normal and aligned-force CMC, adhered to a flat substrate}

\begin{figure}
    \centering
    \includegraphics[width=\linewidth]{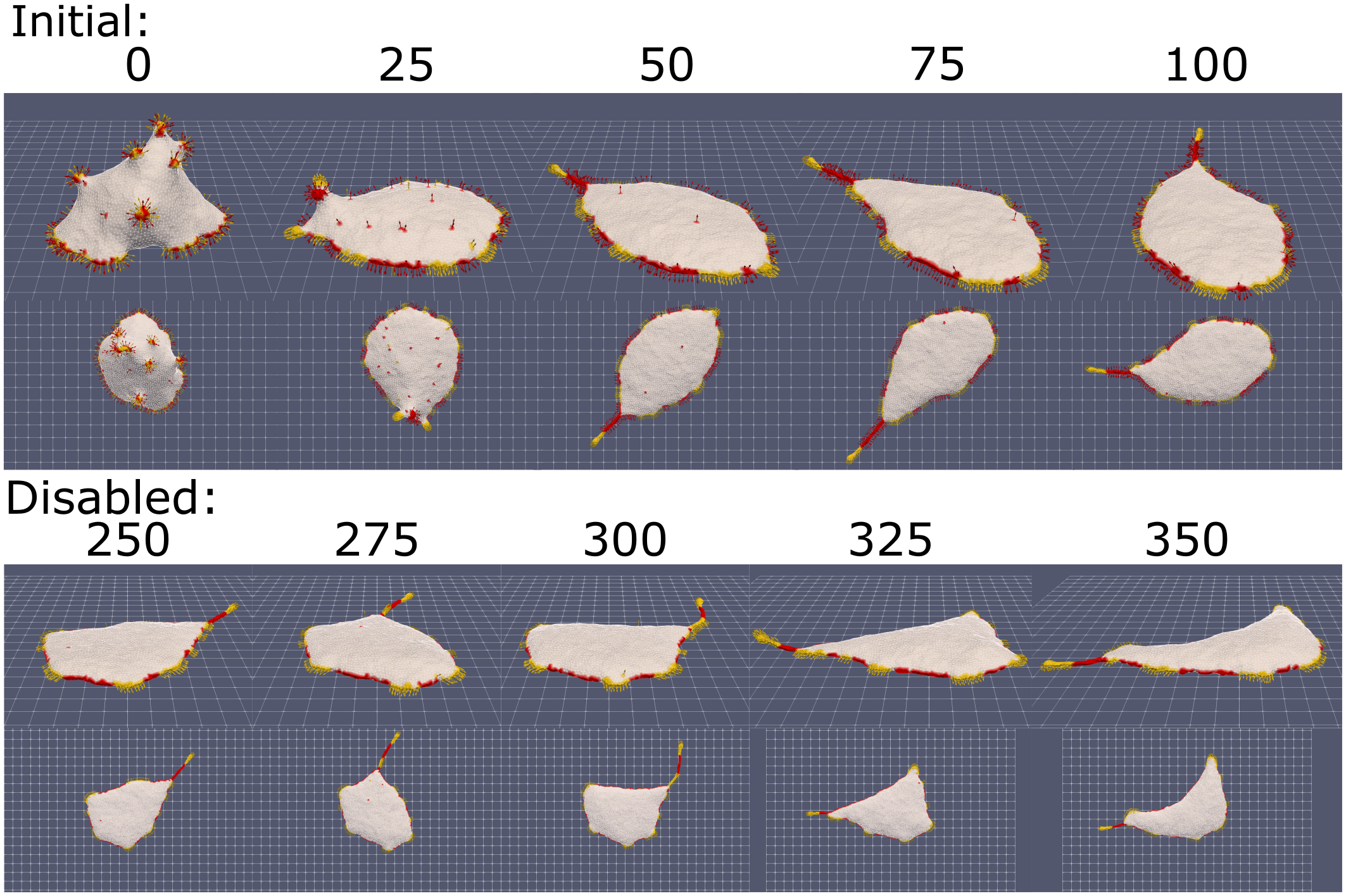}
    \caption{Overview of an adhered vesicle with a mixture of aligned-force (yellow) and normal-force (red) CMC, which have exclusive binding interactions between them (see Fig.8 in the main text). At time $t=250$ the force is disabled for the normal-force CMCs, leaving only the aligned-force CMCs active. The original simulation is given on the top (times $0-100$), and the simulation after the normal-force has been disabled is at the bottom.}
    \label{fig:6_adhesive_disable}
\end{figure}

In Fig.\ref{fig:6_adhesive_disable} we show the dynamics of the vesicle that contains the mixture of aligned-force (yellow) and normal-force (red) CMC, which have exclusive binding interactions between them (see Fig.8 in the main text). At time $t=250$ we turned off the normal-force CMC, keeping only the aligned-force CMC active. We find that the adhered area shape changes, with the rim regions that contain the curved passive (red) CMC retract into the vesicle, while the aligned-force regions protrude more prominently along the adhered rim. 

\section*{Movies}

\begin{itemize}

\item{\textbf{\href{https://weizmann.box.com/s/m4ga8726esr5eik6a78lvn7rztx22eh4}{Movie-S1}}
Aligned-force simulation of the formation of filopodia-like tubular protrusions (corresponding to Fig.7B), with parameters $\kappa=28.5,f=0.2,w=2,C_0=0.4,\rho=20\%,s=0.5,r=30$}

\item{\textbf{\href{https://weizmann.box.com/s/3xwo3lafb0z0hej8r0s20ktp53ovekdg}{Movie-S2}} 
Normal force simulation, in the regime of tubes shapes (corresponding to Fig.7B), with parameters $\kappa=28.5,f=0.5,w=2,C_0=0.1,\rho=20\%$}

\item{\textbf{\href{https://weizmann.box.com/s/172op3owke8y7sqhsva1biti9kivjo0f}{Movie-S3} }Adhered, universal-binding between normal-force CMCs (red) and aligned-force CMCs (yellow), corresponding to Fig.8A. Parameters used: $\kappa=28.5,f=0.5,w=2,w_{ad}=0.25,C_0=0.8,\rho_n=10\%,\rho_a=10\%,s=1,r=15$}

\item{\textbf{\href{https://weizmann.box.com/s/2vecpkv97j12uqfe7xtnetuj7ka9bmbx}{Movie-S4}}
Adhered, exclusive-binding between normal-force CMCs (red) and aligned-force CMCs (yellow), corresponding to Fig.8A. Parameters used: $\kappa=28.5,f=0.5,w=2,w_{ad}=0.25,C_0=0.8,\rho_n=10\%,\rho_a=10\%,s=1,r=15$}

\item{\textbf{\href{https://drive.google.com/open?id=1NZvMIs10UujEyYoKbGJqjYYpH7Dyo7zC&authuser=0&usp=drive_link}{Movie-S5}}. The 3D movie of the cell in Figure 9A}

\item{\textbf{\href{https://weizmann.box.com/s/ugguocnin2wbyeeia21fm0awsweax0jy}{Movie-S6}}. The movie of the XY and XZ section for Figure 9E}

\item{\textbf{\href{https://weizmann.box.com/s/ivn1dyp645qcw2wkunvulmyfiuvfy31r}{Movie-S7}}. The movie of the XY and XZ section for Figure 9F}

\item{\textbf{\href{https://weizmann.box.com/s/pyokxmgqoa3vfemxmteoop9a2xfxyqh3}{Movie-S8}}. The movie of the XY and XZ section for Figure 9G}

\label{movies}
\end{itemize}


\bibliography{references.bib}

\section*{}

\begin{table}
    \centering
    \begin{tabular}{|c|c|c|c|c|c|c|c|}
        \hline
         parameter & units & Fig.\ref{fig:3_analytical_line} & Fig.\ref{fig:4_undulation_comparison} & Fig.\ref{fig:5_Coral_exclusive_phase_diagram} & Fig.\ref{fig:6_adhesive_disable} & movie 1,2 & movie 3,4 \\
         \hline
         $f$ & $K_B T / \ell_{min}$ & $0-1.2$ & $0.5,0$ & $0.5$ & $0.5$ & $0.2$, $0.5$ & $0.5$ \\
         $w$ & $K_B T$ & $0-0.48$ & $2$ & $2$ & $2$ & $2$ & $2$ \\
         $\kappa$ & $K_B T$ &   $20$ & $28.5$ & $28.5$ & $28.5$ & $28.5$ & $28.5$\\
         $\rho$ & 1 & $10\%$ & $10\%,10\%$ & $10\%,10\%$ & $10\%,10\%$ & $20\%$  & $10\%,10\%$ \\
         $C_0$ & $1/\ell_{min}$ & $1$ & -$0.6-0$, $0.8$ & -$0.75-0$, $0.8$ & $0.8$ & $0.4$,$0.1$ & $0.8$\\
         \hline
    \end{tabular}
    \caption{The values of the model parameters used in the simulations in the different figures. The energy units are $K_B T=1$, which define the scale of $f,w,\kappa$, and the length units are $\ell_\text{min}=1$, which define the scale of the vertex lattice, the force, and spontaneous curvature.}
    \label{tab:simulation_parameters}
\end{table}

